\pdfoutput=1
\documentclass[seceq]{ptptex}

\usepackage{graphicx}

\notypesetlogo                       

\title{
GeV Photon Beams for Nuclear/Particle Physics
}

\author{
Norihito \textsc{Muramatsu} for the LEPS Collaboration%
}

\inst{
\small Research Center for Nuclear Physics, Osaka University, Ibaraki, Osaka, 567-0047, Japan
}

\abst{
Production of a GeV photon beam by laser backward-Compton scattering has been playing an important 
role as a tool for nuclear and particle physics experiments. Its production techniques are now 
established at electron storage rings, which are increasing worldwide. A typical photon intensity
has reached $\sim$10$^6$ sec$^{-1}$. In the present article, the LEPS beamline facility at SPring-8 
is mainly described with an overview of experimental applications, for the purpose to summarize the 
GeV photon beam production. Finally, possible future upgrades are discussed with new developments 
of laser injection.
}

\begin{document}

\maketitle

\section{Application to Nuclear/Particle Physics} \label{sec-intro}

  A GeV photon beam is an important experimental tool to investigate hadron structures and reactions. 
A photon couples to a quark and anti-quark pair with a good approximation of the vector meson dominance 
model \cite{vdm}, and interacts with a nucleon or nuclear target producing a combination of mesons and 
a baryon. Measurements 
of differential cross sections and spin observables for such hadron photoproduction not only provide 
information about coupling constants among hadrons in the s-, t-, and u-channel diagrams but also may 
reveal unconventional contributions near the production thresholds, various N$^*$ resonances arising 
from rich quark configurations, and exotic hadron structures including a strong meson-baryon correlation 
and a multi-quark state. In general, a photon beam is clean without hadron contaminations, while 
a pion or kaon beam, which is widely used in hadron experiments, contains other types of mesons and 
baryons produced simultaneously at a second-beam production target. Most significant contamination 
in the photon beam only comes from pair-created electrons and positrons, but they can be separated by 
a sweeping magnetic field, a charged particle veto counter, and an Aerogel ${\rm \check{C}}$erenkov 
counter. A photon beam is characterized by the orientation of its electric or magnetic field, and
it is possible to make it polarized linearly, circularly, or elliptically. Such 
a polarized beam enables to inquire into the spin-parity of an intermediate state in a hadron reaction.
A disadvantage of using a photon beam is a low reaction rate, compared with the case of a hadron beam. 
On the other hand, this feature as well increases the relative rate of a photon-nucleon reaction deeply 
inside a nucleus, and provides an advantage for a study of mesic nuclei, which are interesting to 
investigate a restoration of chiral symmetry breaking in high density materials.

  The GeV photon beam is usually produced from two methods: backward Compton scattering of laser light 
from a high energy electron beam and Bremsstrahlung radiation from a high energy electron beam at a thin 
radiator. The former method has been adopted in the LEPS experiment at SPring-8 \cite{leps}, the GRAAL 
experiment at the European Synchrotron Radiation Facility (ESRF) \cite{graal}, and the LEGS experiment at 
the National Synchrotron Light Source (NSLS) of Brookhaven National Laboratory (BNL) \cite{legs1,legs2}, 
whereas the latter method has been conducted in the CLAS experiment at Jefferson Laboratory (JLab)
\cite{jlab}, the SAPHIR, Crystal Barrel/TAPS, and GDH experiments at Bonn-ELSA \cite{saphir,cbelsa,gdhelsa}, 
and the Crystal Ball/TAPS and GDH experiments at Mainz-MAMI \cite{tapsmami,gdhmami}. While the maximum 
photon energy is limited up to the Compton edge in the method by the backward Compton scattering, an energy 
spectrum of the backscattering photons is relatively flat retaining an order of photon intensity over a wide 
energy range. In contrast, the photon intensity of a Bremsstrahlung beam rapidly decreases with an increasing 
photon energy. Since many of the lasers supply a linearly polarized light, 
the backscattering photon beam possesses high 
polarization degree after the head-on collisions with high energy electrons. The photon beam polarization 
can be easily handled in a desired direction by an optical control of the laser polarization. 
A laser-backscattering facility can be co-operated with a high current electron storage ring without 
extracting an electron beam to an experimental area equipped with a beam dump, and opportunities to 
construct such facilities are increasing because of new demands for synchrotron radiation storage rings, 
where material, chemical, and biological sciences are intensively studied.

  The maximum energy ($k_{max}$) of backscattering photons is kinematically determined with the following 
equation under a condition of head-on collision:
\begin{equation}
      k_{max} = \frac{(E_e + P_e) k_{laser}}{E_e - P_e + 2 k_{laser}}
         \simeq \frac{4 {E_e}^2 k_{laser}}{{m_e}^2 + 4 E_e k_{laser}} ,
\end{equation}
where $E_e$, $P_e$, and $m_e$ indicate the energy, momentum, and mass of an electron in a storage ring, 
respectively, and $k_{laser}$ corresponds to the energy of laser light. Typically, a GeV photon beam is
produced by magnifying an ultraviolet (UV) laser energy of several eV using several GeV electrons. In 
this energy scale, the maximum photon energy becomes roughly proportional to the laser energy and a square
of the electron beam energy. Therefore, the electron energy of a storage ring is an important factor to 
determine an available photon energy range, which prescribes for physics interests. The energy of 
backscattering photons kinematically depends on a scattering angle relative to the electron beam direction, 
but the scattering cone angle of the GeV photons is limited to $\sim$1/$\gamma$ radian by a Lorentz factor 
of the electron beam. The scattering photons fly back nearly on the injected laser axis, and a photon beam 
size is well collimated to a few cm diameter, where a small fixed target is usable. Up to now, UV and deep 
UV lasers with the wavelengths of $\sim$350 nm and $\sim$260 nm, respectively, have provided reasonable
output powers at 1$\sim$9 W, which results in the tagged photon beam intensity of 0.1$\sim$4$\times$10$^6$ 
sec$^{-1}$ with an electron beam current of 100$\sim$350 mA. Nearly 100\% of the laser polarization is 
transferred to the backscattering photon around the Compton edge because an electron spin-flip amplitude 
disappears. Since a beam electron loses a large fraction of energy giving a few GeV to a photon, a recoil 
electron deviates from the storage beam trajectory in a magnetic field. A bending magnet of the storage 
ring is usually utilized for tagging such a recoil electron at a segmented detector, and its momentum is 
measured based on a displacement of track position in order to determine a scattering photon energy. 
Practically, high recoil momentum electrons or low energy photons cannot be tagged because of the 
geometrical acceptance of a tagging detector.

\begin{table}[t]
 \caption{Properties related to the productions of the laser-backscattering photon beam at the LEPS, GRAAL,
          and LEGS experiments. `Tagger Threshold' and `Tagged Intensity' represent the lowest energy of 
          tagged photons and a photon beam intensity counted at a tagging detector, respectively. The LEPS
          experiment uses all solid-state lasers, each of which provides a single wavelength at 355 nm (UV =
          ultraviolet) or 266 nm (DUV = deep ultraviolet), whereas the other experiments operate an Argon-gas
          laser, which gives either of multiline wavelengths around 351 nm (MLUV = multiline ultraviolet) or
          a single line wavelength from 302 nm (DUV) to 514 nm (Green). Photon energy resolutions of all the
          experiments are determined not by a position resolution of a tagging detector but by an electron 
          beam emittance.}
 \label{tab:bcsfacility}
 \centering
 \begin{tabular}{|l|ccc|}
 \hline
 Facility Name         &        LEPS         &        GRAAL           &         LEGS           \\
 \hline\hline
 Storage Ring          &      SPring-8       &        ESRF            &       NSLS (BNL)       \\
 e$^-$ Beam Energy     &      7.975 GeV      &      6.03 GeV          &  Phase-I  : 2.584 GeV  \\
                       &                     &                        &  Phase-II : 2.800 GeV  \\
 e$^-$ Beam Current    &       100 mA        &    150$\sim$200 mA     &    300$\sim$350 mA     \\
 Laser Wavelength      &     UV : 355 nm     &    Green : 514 nm      &  MLUV:333,351,364 nm   \\
                       &    DUV : 266 nm     &  MLUV:333,351,364 nm   &    DUV : 302 nm        \\
 Laser Power           &     UV : 8 W        &     Green : 10 W       &   MLUV : $\leq$9 W     \\
                       &    DUV : 1 W        &     MLUV  :  8 W       &    DUV : 3 W           \\
 Maximum E$_\gamma$    &     UV : 2.4 GeV    &   Green : 1.1  GeV     & Phase-I:0.33 GeV(MLUV) \\
                       &    DUV : 2.9 GeV    &   MLUV  : 1.53 GeV     & Phase-II:0.42 GeV(DUV) \\
 Tagger Threshold      &        1.5 GeV      &      0.55 GeV          &    $\sim$0.2 GeV       \\
 Tagging Method        &  Bending Magnet +   & Bending Magnet +       & Bending Magnet +       \\
                       &  Silicon Strips or  & Silicon Strips +       & 3 External Magnets +   \\
                       &  Scint. Fibers +    & Trigger Counters       & Plastic Scint. Bars    \\
                       &  Trigger Counters   & (Internal Tagging)     &                        \\
 Tagged Intensity & UV:1$\sim$3$\times$10$^6$ $\gamma$/sec & Green:2$\times$10$^6$ $\gamma$/sec & MLUV:4$\times$10$^6$ $\gamma$/sec \\
                  & DUV:1$\sim$3$\times$10$^5$ $\gamma$/sec & MLUV:1$\sim$2$\times$10$^6$ $\gamma$/sec &                            \\
 E$_\gamma$ Resolution &  12 MeV ($\sigma$)  &    16 MeV (FWHM)       &    5.4 MeV (FWHM)      \\
 \hline
 \end{tabular}
\end{table}

  The facilities which have been constructed to produce a laser-backscattering photon beam at the 
GeV-scale complementarily cover different photon energy ranges depending on the electron accelerator 
energy \cite{bcscomp}. Properties of the existing facilities and the produced photon beams 
are summarized in Table~\ref{tab:bcsfacility}. The highest photon energy region 
is currently covered by the LEPS experiment at SPring-8, which provides a 7.975 GeV electron beam. 
A photon beam up to 2.4 GeV is produced by using a UV laser light with a wavelength of 355 nm. Only 
the energy region above 1.5 GeV is tagged with a rate of 1$\sim$3$\times$10$^6$ 
sec$^{-1}$, which has been improved by a simultaneous injection of two lasers as discussed later.
The covered energy range exceeds the $K\bar{K}N$ threshold of 1.5 GeV, so that photoproductions of 
a $\phi$ meson, hyperon excitations, and exotic baryons including strangeness have been intensively 
studied. The LEPS experiment has further extended the maximum photon energy up to 
2.9 GeV with a tagging rate of 1$\sim$3$\times$10$^5$ sec$^{-1}$ by introducing a deep UV laser with 
a wavelength of 266 nm. This energy upgrade has enabled to study photoproductions of hyperon excitations 
with a K$^*$ meson exceeding $K^*\bar{K}N$ threshold at 2.4 GeV, and to achieve a recoilless condition of 
$\rho$ and $\omega$ mesons at 2.75 GeV in their photoproductions inside a nucleus. The GRAAL experiment 
at ESRF, where a 6.03 GeV electron beam is stored, covers a photon energy range from 0.55 GeV to 1.53 GeV. 
Photons from an 8 W Ar laser, whose wavelength is multiline around 351 nm, are scattered off the electron 
beam of 150$\sim$200 mA. As a result, a backscattering photon intensity reaches 1$\sim$2$\times$10$^6$ 
sec$^{-1}$, which is counted by a tagging detector placed inside the storage ring (internal tagging).
Photoproductions of light pseudoscalar mesons and ground state hyperons off nucleons have been analyzed 
to measure differential cross sections and photon beam asymmetries for the purpose to study photoproduction 
mechanisms at low energies and N$^*$ resonances in the s-channel \cite{phygraal}. At the BNL-LEGS experiment, 
Ar laser lights with wavelengths from 302 nm to 514 nm are injected into the 2.5$\sim$2.8 GeV electron storage 
ring at NSLS, producing 0.421 GeV photons at most. Since the electron beam current reaches 300$\sim$350 mA,
a typical tagged photon intensity has been observed at 4$\times$10$^6$ sec$^{-1}$ with a multiline UV
laser. Pion photoproduction and Compton scattering off nucleons have been investigated at a $\Delta$ 
resonance region in order to study multipole amplitudes and proton polarizabilities \cite{legs1}. 
A polarized HD target has been also developed for a double polarization experiment to proceed 
further analyses about polarizabilities and GDH sum rule \cite{legs2}.

  As discussed above, a GeV photon beam by backward Compton scattering is useful for various hadron and
nuclear physics programs depending on its energy range, whereas the production techniques are common in
many parts. In Section~\ref{sec-lepsfac}, we discuss in detail about the production of a GeV photon beam 
at the LEPS experiment, as an example, and overview its physics results.

\section{LEPS Facility at SPring-8} \label{sec-lepsfac}

\subsection{GeV Photon Beam Production at SPring-8}

  SPring-8 \cite{sp8sr} is a third generation facility of synchrotron radiation light sources, and provides 
the highest electron energy among them. 62 beamlines are available on the electron storage ring, whose 
circumference is 1,436 m. There are forty of 7.8 m-long straight sections and four of 30 m-long straight 
sections between 30 m-long cells, which consist of two bending magnets, 10 quadrupole magnets, and 7 sextupole 
magnets. The LEPS facility utilizes one of the 7.8 m-long straight sections to collide laser photons with 
7.975 GeV electrons. A recoil electron is tagged just downstream of a bending magnet, located at the end of 
the straight section, in order to measure the energy of a scattering photon. Electrons with a current of 100 
mA are bunched to 13 psec length by a 508.58 MHz RF frequency, which corresponds to the minimum interval of
1.966 nsec. Several variations exist for a filling pattern of electrons by making a few hundreds of bunches
isotropic on the circumference or by filling several single bunches and long bunch trains. In the case that 
the intensity of a backscattering photon beam reaches 10$^6$ sec$^{-1}$, the laser-electron collision occurs
at about five electron bunches on a circumference. An electron beam size ($\sigma$) at the 7.8 m-long straight 
section is 295 $\mu$m and 12 $\mu$m on averages in the horizontal and vertical directions, respectively, 
so that the injected laser light must be well focused to these levels in size.

\begin{figure}[t]
 \centering
 \includegraphics[scale=0.6]{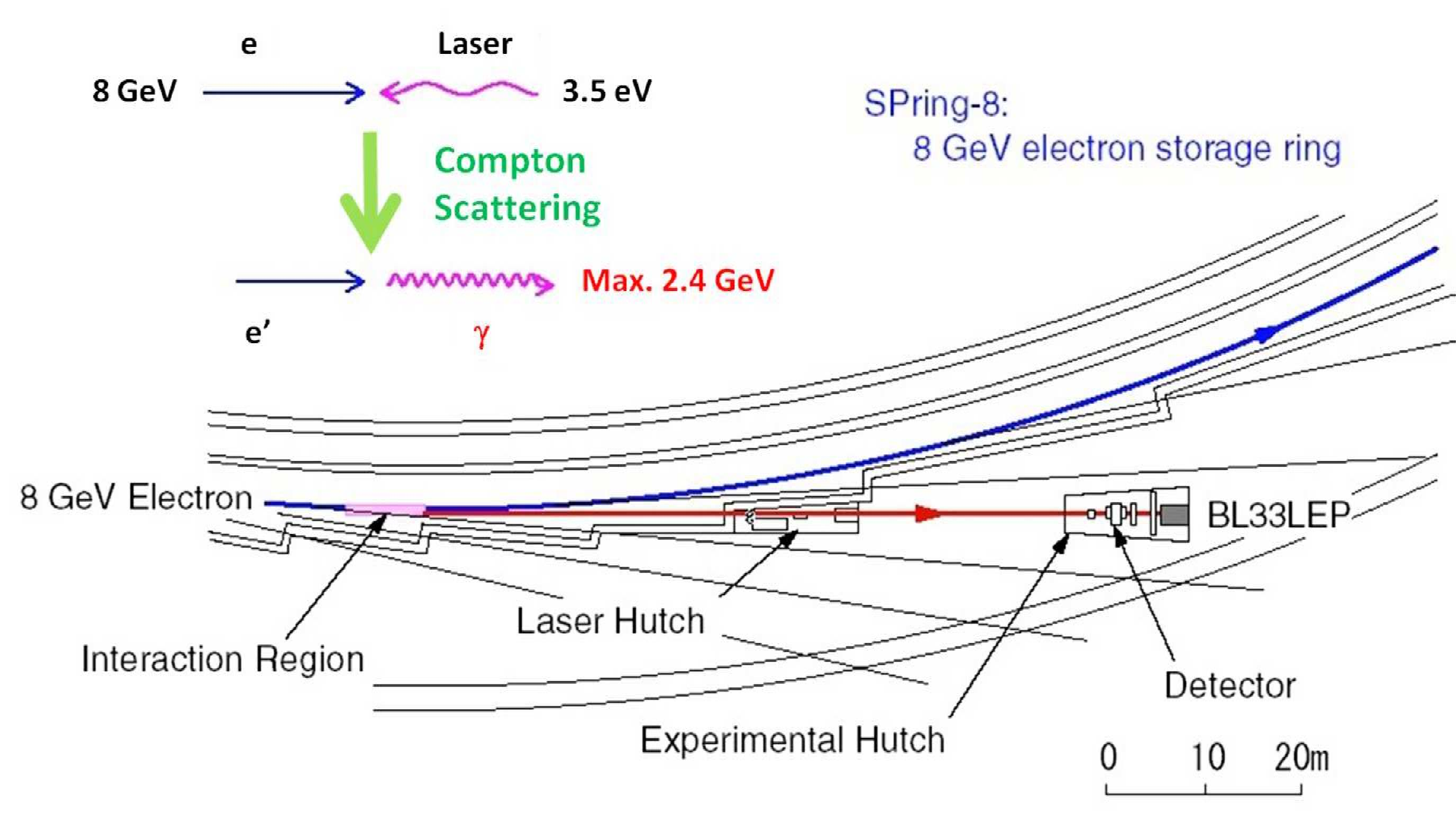}
 \caption{A plan view of the LEPS facility (BL33LEP), where a laser-backscattering photon beam is produced.
          A laser light is injected from `Laser Hutch' into SPring-8. Photons are backwardly scattered at
          `Interaction Region' by 8 GeV electrons with energy magnification up to a few GeV, and delivered 
          to `Experimental Hutch', where a target material and detectors are set up.}
 \label{fig:bl33lep}
\end{figure}

  Figure~\ref{fig:bl33lep} shows a plan view of the laser-backscattering beamline (BL33LEP) at SPring-8. 
A laser injection system is placed inside a radiation shield hutch (`laser hutch'), which is located besides 
the storage ring tunnel surrounded by 1 m-thick concrete walls. The laser hutch is made of iron and lead walls
with a floor size of 2$\sim$3 m width $\times$ 12 m length, and its entrance and cable ducts are interlocked 
with a $\gamma$- and x-ray beam shutter. Inside the hutch, lasers and optical components, which will be discussed 
in Subsection~\ref{ss-inj}, are set up on a steel honeycomb surface plate with a size of 1.2 m $\times$ 3.6 m. 
The whole injection system is protected against dust by assembling a clean booth on the surface plate. A laser 
light travels about 37 m from the laser hutch to a scattering point inside beamline chambers, which are directly 
connected to the storage ring with a ultrahigh vacuum of 10$^{-8}$ Pa. A part of laser photons collide with high 
energy electrons at the 7.8 m-long straight section of the storage ring. The scattering photons with GeV energies 
are directed backwardly on the injected laser path, and delivered through a medium vacuum pipe ($\sim$1 Pa) 
to an experimental hutch, which is also made of radiation shields with a floor size of 5$\sim$6 m $\times$ 12 m.
The location of the experimental hutch is about 70 m downstream of the straight section. A detector assembly 
and a target material are set up inside this hutch in order to study photoreactions on interests. Further 
details of a laser injection, a beamline structure, and properties of the produced photon beam are discussed 
in Subsections~\ref{ss-inj}, \ref{ss-bl}, and \ref{ss-prop}, respectively, along with a description of 
the LEPS experiment in Subsections~\ref{ss-det} and \ref{ss-phy}.

\subsection{Laser Injection} \label{ss-inj}

\begin{table}[t]
 \caption{Properties of four lasers, which have been used at the LEPS experiment. `Sabre' series of Coherent
          Inc. use Argon-gas tube as a cavity, which must be cooled by a large water flow supplied from
          the SPring-8 facility. The other lasers are operated by a diode laser, which works as a seed, and
          then obtain a UV or deep UV wavelength by doubling or tripling the seed frequency with a nonlinear 
          crystal. Power consumption of these lasers are low, so that it is enough to prepare internal water 
          circulation between a laser head and a controller or only air cooling. The `1/e$^2$ diameter' below
          corresponds to a 2$\sigma$ diameter of a Gaussian beam.}
 \label{tab:lepslaser}
 \centering
 \begin{tabular}{|l|cccc|}
   \hline
   Laser name         &    Innova Sabre      & Paladin 355-8000  &    Innova Sabre      &   Chicon      \\
                      &       DBW25/7        &                   &      Moto Fred       &               \\
                      &   (Coherent Inc.)    & (Coherent Inc.)   &   (Coherent Inc.)    & (Oxide Corp.) \\
   \hline\hline
   wavelength         &    multiline UV      &    355 nm         &     257.2 nm         &    266 nm     \\
                      & 333.4$\sim$363.8 nm  &                   &                      &               \\
   emission method    &     Ar-gas tube      &   diode laser +   &     Ar-gas tube +    & diode laser + \\
                      &                      &   nonlin. cryst.  &     nonlin. cryst.   & nonlin. cryst.\\
                      &                      &    (mode-lock)    &                      &               \\
   emission frequency &        CW            &    80 MHz         &        CW            &     CW        \\
                      &                      &   (quasi-CW)      &                      &               \\
   UV output power    &        7W            &     8 W           &       1 W            &    1 W        \\
   1/e$^2$ diameter   &      1.7 mm          &    1.0 mm         &   0.6$\sim$0.9 mm    &    3.0 mm     \\
   divergence         &     0.31 mrad        &   0.55 mrad       &  0.5$\sim$0.85 mrad  &    0.4 mrad   \\
   power consumption  &       10 kW          &     350 W         &      10 kW           &   300 W       \\
   cooling method     &   water cooling      &  internal water   &   water cooling      &  air cooling  \\
                      &   $\sim$25 l/min     &   circulation     &   $\sim$25 l/min     &               \\
   \hline
 \end{tabular}
\end{table}

  Table~\ref{tab:lepslaser} shows properties of the UV and deep UV lasers used at the LEPS experiment so far. 
From the start of the LEPS experiment in 1999, a multiline Argon-gas tube laser, which has been widely adopted 
for the production of a laser-backscattering photon beam, was operated for a long time. Main wavelengths of 
this laser were 333.4 nm, 351.1 nm, and 363.8 nm, which make the Compton edges at 2.49 GeV, 2.40 GeV, and 2.34 
GeV, respectively. The total power of UV output was about 7 W at the maximum tube current, resulting in 
a tagged photon beam intensity of 
$\sim$10$^6$ sec$^{-1}$. There are some disadvantages in the use of the Argon laser: a necessity to prepare 
infrastructures supplying a large electric power and a huge cooling water flow, and a periodical replacement 
of the Argon-gas tube for keeping a reasonable output power. Hence, a new solid-state mode-lock laser, whose 
wavelength is 355 nm, has been introduced from 2006. Its output power is fixed at 8 W with a beam diameter
of 1.0 mm, which is defined at a 1/e$^2$ contour. The obtained Compton edge is mostly unchanged at 2.39 GeV, 
and the tagged photon beam intensity reaches $\sim$1.5$\times$10$^6$ sec$^{-1}$ thanks to a slight increase 
of the laser power, a net gain by a long-term power stability, optimizations of laser injection optics, and
a start of top-up operation at SPring-8. The electric power consumption of this laser is only 350 W with 
internal water circulation, which is primarily air-cooled. 
A laser pointing and a UV output power were reasonably stable during 
a few year operation, while it was sometimes necessary to shift a damaged laser spot to a new one on a nonlinear 
crystal, which makes a seed laser wavelength one third by a third harmonic generation (THG). 
Although the laser emission is pulsed with a frequency of 80 MHz in a quasi-CW mode, laser photons, which are
focused at the straight section making a few m-long beam waist, sufficiently collide with the electron beam,
which is bunched in each 2 nsec corresponding to a running distance of 0.6 m. Because of the reduction of 
electric power consumption and cooling water flow, it becomes possible to operate multiple lasers simultaneously, 
as discussed later.

  As mentioned in Section~\ref{sec-intro}, the shorter wavelength of an injecting laser provides the higher 
maximum energy of a laser-backscattering photon beam. The LEPS experiment had previously used an Argon 
laser-based deep UV laser with a wavelength of 257.2 nm, which raises the Compton edge up to 2.96 GeV. 
The deep UV wavelength is achieved 
by a frequency doubling, arising from a second harmonic generation (SHG) at a Barium Borate (BBO) crystal with
an input of the green line (514.4 nm). The deep UV output was typically 1 W with a CW emission. A usable time 
range of a laser spot on the BBO crystal was 1$\sim$2 week, so that the BBO crystal had to be frequently shifted. 
In addition, electric power consumption and cooling water flow are huge because the same Argon-gas tube system 
as the above-mentioned UV laser is utilized as a seed laser. These problems have been recently solved by 
introducing a new solid 
state laser, where a laser light with a wavelength of 266 nm is produced similarly by a SHG at a BBO crystal from 
a green light output of another SHG using a Lithium Triborate (LBO) crystal and an infrared (IR) diode laser. 
Although the maximum energy of the backscattering photon beam slightly drops to 2.89 GeV, a stable output of 1 W 
with a CW emission lasts more than a half year because of a significant improvement of a BBO crystal purity, which 
as well increases a SHG efficiency up to $\sim$20\%. As a result, a tagged photon intensity of $\sim$1.5$\times$10$^5$ 
sec$^{-1}$ have been obtained constantly. The electric power consumption is reduced to 300 W, and a heat generation 
becomes small enough enabling an air cooling. Only a control of environmental temperature for the laser head has 
been tightly necessary within 1 $^\circ$C. Generally, a high power laser with a deep UV wavelength tends to cause 
unrecoverable damages at optical components of a laser injection system, but the new deep UV laser minimizes this 
kind of damages thanks to a large beam diameter of 3 mm.

\begin{figure}[t]
 \centering
 \includegraphics[scale=0.55]{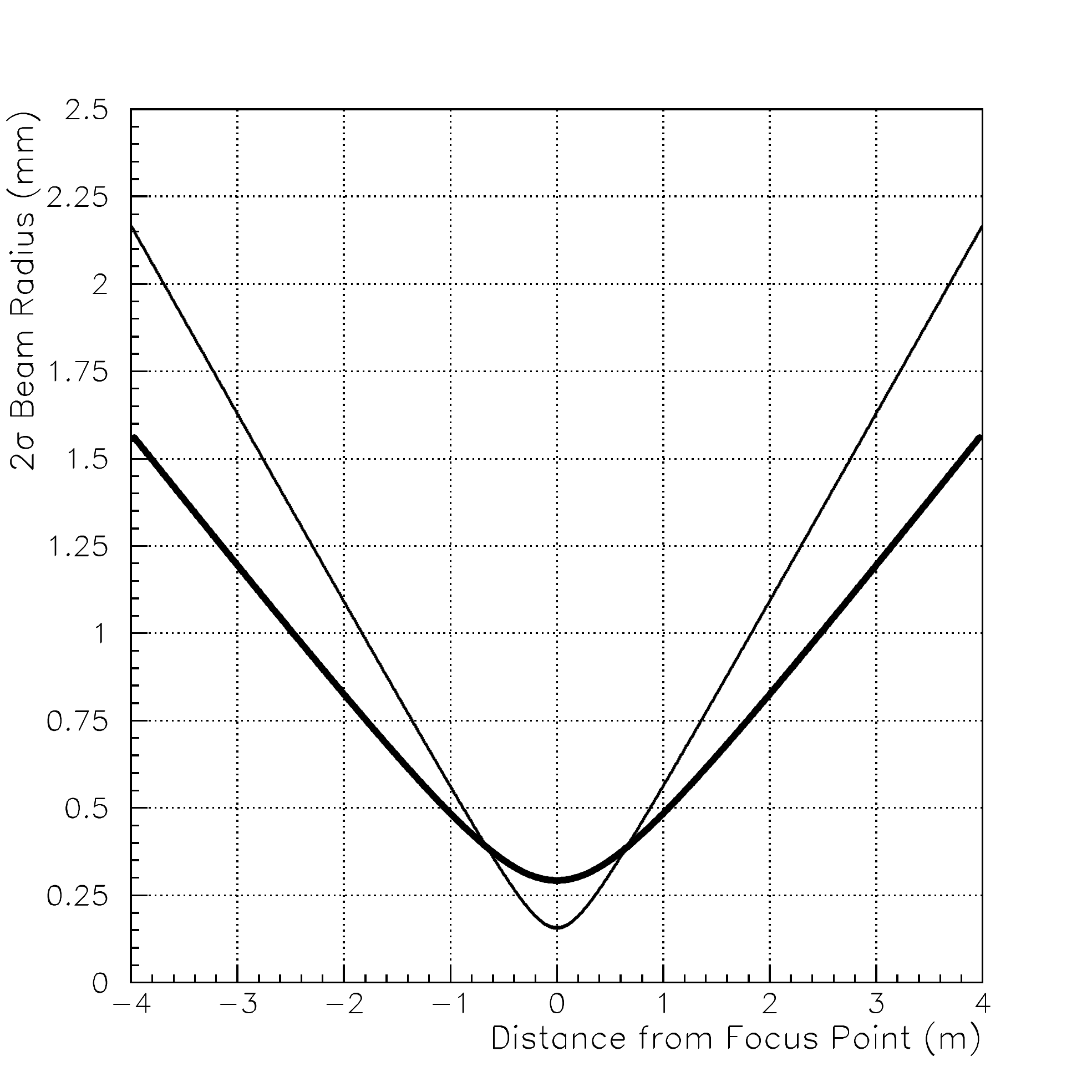}
 \caption{2$\sigma$ or 1/e$^2$ radii of the laser beams used at the LEPS experiment as a function of a distance 
          from a focus point. The radii are calculated for the region around the focus point, based on an ideal
          Gaussian beam propagation by Eqn.~\ref{eq:laserprop}. 
          The thick line shows a propagation of the UV laser with $\lambda$$=$355 nm, while the thin line 
          indicates the case of deep UV laser with $\lambda$$=$266 nm.}
 \label{fig:focus}
\end{figure}

  The laser light must be focused to several hundred $\mu$m diameter, corresponding to the electron beam size, 
at the 37 m downstream straight section of the storage ring. In Gaussian beam optics, it is known that a 1/e$^2$ 
radius (r [mm]), which is the same as 2$\sigma$, is described by the following propagation equation:
\begin{equation}
      r = w_0 \times \sqrt{1+\left(\frac{\lambda \times |z|}{\pi \times {w_0}^2 \times 10^3}\right)^2} \;\; , \label{eq:laserprop}
\end{equation}
where $w_0$ [mm] is a $1/e^2$ radius of the beam waist at a focus point, $\lambda$ [nm] is a wavelength of 
the laser, and z [m] is a distance from the focus point. This equation tells that the laser beam diameter has 
to be expanded at a starting point of propagation up to a design value, which is calculated from a desired beam 
waist size and a necessary distance to the focus point. For the LEPS experiment, magnification factors 
of the expansion are set to 28.6 and 13.3 for the UV laser ($\lambda$ $=$ 355 nm) with a 1.0 mm diameter and 
the deep UV laser ($\lambda$ $=$ 266 nm) with a 3.0 mm diameter, respectively, so that the expanded beam diameters
should become 29 mm and 40 mm at the laser hutch. In the ideal case without taking into account beam qualities
(M$^2$) and diffractions at optics, these beam expansions cause 1/e$^2$ radii around focus points to behave
as shown in Fig.~\ref{fig:focus}. The magnification factor for the UV laser is determined 
to be a slightly lower value so that the same injection optics can be used for another UV laser, which possesses 
a larger diameter of 1.4 mm and will be used in future because of a higher power output, as discussed in 
Section~\ref{sec-future}.

  An optical expansion of beam diameter is performed by a Galilean telescope, called a beam expander, which 
is placed $\sim$0.5 m downstream of the laser. The beam expander consists of a set of concave lenses to 
increase the input beam diameter and a convex lens to make a focus of the expanded laser output. The diameter 
of the output convex lens is designed to be 40 mm, and a tail part of the laser beam, whose diameter becomes 
more than 40 mm, is truncated by an internal structure of the beam expander to remove side lobes. The optical 
lenses of the beam expander are made of UV grade synthetic fused silica, which minimizes hydroxyl (OH) 
concentrations to reduce generations of color centers by a high power laser. Surfaces of the lenses are coated 
by the material which only passes a light with a desired wavelength (anti-reflection or AR coating), and the 
expanders are, therefore, separately produced for the wavelengths of 355 nm and 266 nm. F numbers of the concave 
and convex lenses, representing {\it a lens focus / a lens diameter}, are determined to 5 for both of UV and 
deep UV beam expanders, so that total lengths of the expander bodies reach a few tens of cm. On the beam expanders, 
a micrometer to change a distance between the concave lenses and convex lens is mounted for varying a laser 
output diameter and a focal length. A focus position of the laser beam is precisely adjusted with the micrometer 
by monitoring a collision rate of laser photons with electrons.

  In the 10 year experience of the LEPS experiment, the optical lenses of the beam expander were damaged by 
various reasons, and several measures have been taken so far. First, the expander body made of aluminum is 
processed by a hard alumite coating without a black color painting to prevent organic dye vaporization, which 
can soil the lenses. Secondly, glue with silicon oil is avoided at an expander body construction in order to 
eliminate silicon attachment on the lens surface. At the concave lenses, where a laser light is injected with 
a high power density, attachment of dust on the surface drops a transmission rate and a focusing quality, so 
that an inexpensive shield plate with an AR coating is additionally mounted in front of the concave lens and 
nitrogen gas is further blown on the shield plate surface. In the case of using a deep UV laser, unrecoverable 
damage of the AR coating is problematically caused at the concave lens by a long term exposure. This damage is 
basically unavoidable, and a replacement of the concave lenses is necessary. However, introducing the new deep 
UV laser with a diameter of 3 mm has proved that lowering a power density of a laser light is very effective 
to minimize such damage.

  The laser light possesses nearly 100\% linear polarization because of a Brewster window of a cavity or the
birefringence of a nonlinear crystal, and it is transferred to the high energy photon beam at the backward 
Compton scattering. A direction of the linear polarization is controlled by letting the laser beam pass
through a half-wave plate, which generates a phase difference by 
$\lambda$/2 in two perpendicular components of the electromagnetic field vector. A zeroth-order air-gap type
of the half-wave plate, which is produced by overlapping two quartz plates with 
a thickness similar to each other, has been used for thermal compensations under the exposure of a high power 
laser. Previously, a small size half-wave plate, which is commercially available, was used between the laser 
output and the beam expander, but unrecoverable damage was seriously caused at a laser spot and a slight shift 
of the spot position was frequently needed. In order to avoid such cares, a large size half-wave plate, whose 
diameter is 48 mm, has been instead introduced at the downstream of the beam expander, lowering a power 
density. The direction of linear polarization is changed vertically or horizontally by rotating the wave 
plate by 45 degree. In the LEPS experiment, the whole laser injection system is set up inside the interlocked 
laser hutch, so that the rotation of the wave plate is performed remotely through a stepping motor, which is
handled by a pulse motor controller. A series of commands to work this controller are sent from a Labview 
program on a PC via Ethernet and GPIB with a converter. A polarization direction was switched to the other 
typically each half day during physics data taking.

  The expanded laser beam is guided into an entrance port of a vacuum chamber, which is one of beamline chambers
connected to the storage ring, after reflections at two mirrors, called the third and fourth mirrors. These mirrors 
are made of UV grade synthetic fused silica, whose surface is processed by a high reflection (HR) coating for 
the use at an incidence angle of 45 degree. A diameter of the third and fourth mirrors is designed to be 80 mm 
with a surface precision of $\sim$$\lambda$/4. Each of the two mirrors is individually set to a Gimbal holder, 
and mounted on a set of stepping motors, which vary horizontal rotation and vertical elevation angles. By using 
two mirrors, a parallel shift of a beam axis becomes possible in addition to a simple adjustment of a beam direction. 
The HR coating is optimized only for either of the UV or deep UV wavelength, and the mirrors for a desired wavelength 
are chosen to be set on the common holders depending on experimental demands. The stepping motors are remotely 
controlled in the same way as the half-wave plate. Generally, a reflection rate of S-wave, whose polarization is 
perpendicular to a plane of incidence, is slightly higher than that of P-wave, whose polarization is parallel to 
the same plane. Since two additional mirrors for horizontal reflections are fixed inside the vacuum chambers, setting 
the third and fourth mirrors to make vertical reflections cancels a difference of laser beam intensities for vertical 
and horizontal linear polarizations.

\begin{figure}[b]
 \centering
 \includegraphics[scale=0.65]{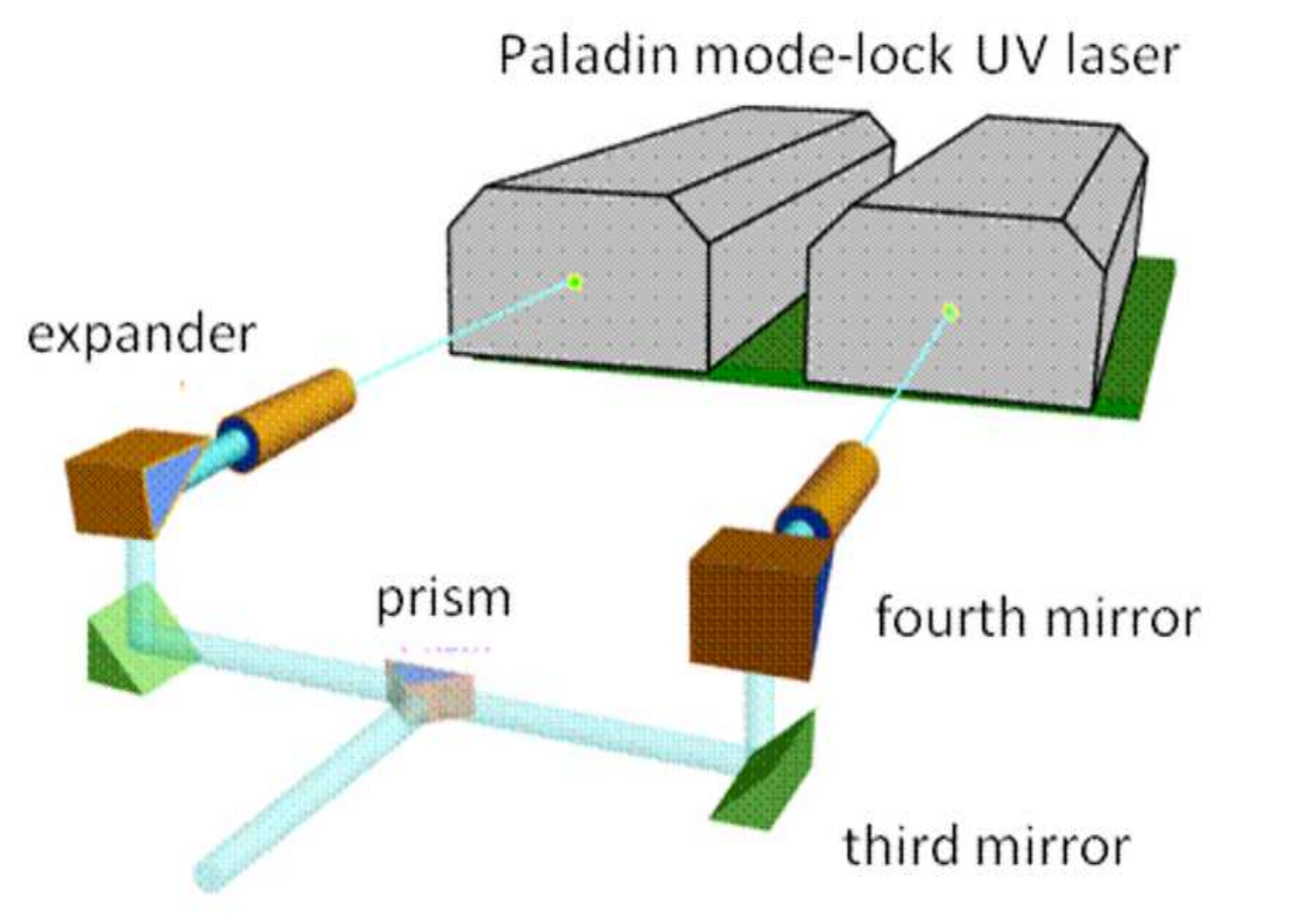}
 \caption{Setup of the simultaneous injection of two mode-lock UV lasers. Laser outputs from the beam expanders
          are reflected at the fourth mirrors (upper mirrors in the figure) and the third mirrors (lower mirrors).
          Large half-wave plates are placed between the beam expander and the fourth mirror during the experiment.
          Two laser beams are merged at the prism toward the beamline chambers. A similar setup with replaced
          optics is also applied to the two laser injection of deep UV lasers.}
 \label{fig:2laser}
\end{figure}

  As described already, solid state lasers with low electric power consumption and low cooling water flow enable 
a simultaneous operation of multiple lasers, which must increase the intensity of a backscattering photon beam 
roughly in proportion to the number of injected lasers. However, an aperture size of beamline limits the number 
of laser beams, whose cross sections are increased by beam expanders. In the case of BL33LEP, only a 40 mm 
$\times$ 40 mm square region is allowed at the entrance port of vacuum chamber for further propagation of 
the lasers to a collision point, so that a simultaneous injection of two lasers is maximally possible. 
Figure~\ref{fig:2laser} shows the setup of two-laser injection at the LEPS experiment. The two laser 
beams are individually expanded by beam expanders, and reflected by the third and fourth mirrors after 
the optical handling by large half-wave plates, similarly as the one laser injection. Then, they are merged to the
same direction next-by-next at a right-angle prism, whose two perpendicular planes are used as reflection mirrors. 
The prism is made of UV grade synthetic fused silica with a HR coating optimized for either wavelength of 355 nm or 
266 nm. The two prisms are replaceable on a holder, which is further mounted on a XYZ stage to adjust a merged beam 
position. Since an intensity of the backscattering photon beam is maximized in a case near the head-on collision of 
laser photons and electrons, the prism stage is adjusted along the merged beam direction so that the two expanded 
beams should be closely adjoined. The stages normal to the merged beam are also aligned to the center axis where a 
true head-on collision must be achieved. The tagged photon intensities have finally reached 2$\sim$3$\times$10$^6$
sec$^{-1}$ and 2$\sim$3$\times$10$^5$ sec$^{-1}$ for the UV and deep UV laser injections, respectively.

\subsection{Beamline Setup} \label{ss-bl}

\begin{figure}[b]
 \centering
 \includegraphics[scale=0.46]{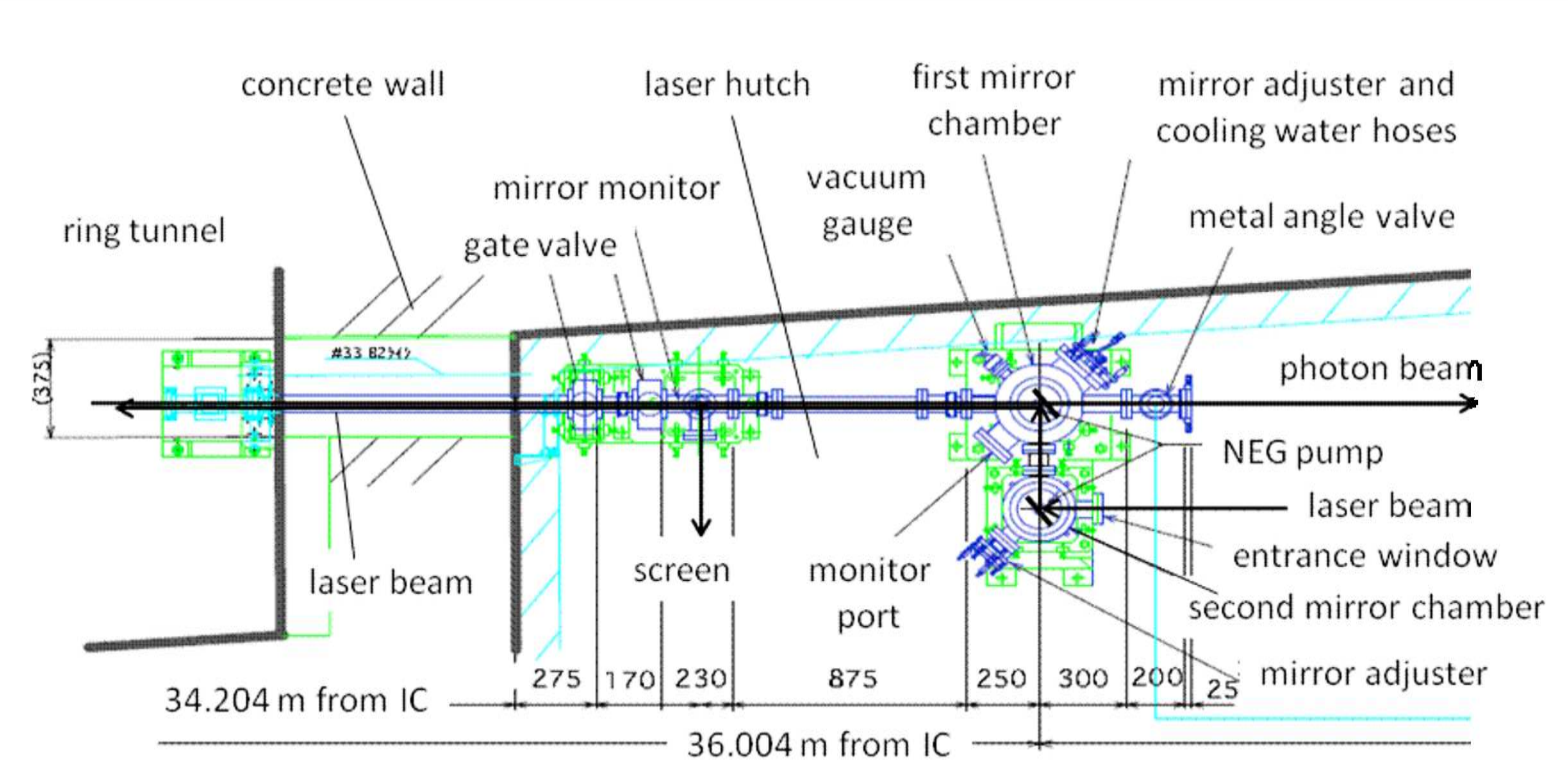}
 \caption{Vacuum chambers related to a laser injection and a backscattering photon beam extraction. A laser beam is 
          injected into a lower-right chamber, and guided by two mirrors toward the storage ring, which is located 
          far in the left-hand side. The backscattering photons travel the same vacuum chambers to the right direction, 
          and exit the chambers at the laser hutch, which is shown in the right part of the figure. A monitor mirror,
          which is set up inside the laser hutch, can be inserted to the laser path near the gateway to the ring
          tunnel anytime when it is desired to extract the laser beam to a screen for monitoring a beam position 
          and measuring a laser power.}
 \label{fig:12mirror}
\end{figure}

  As shown in Fig.~\ref{fig:12mirror}, the laser beam enters a long vacuum beamline at the laser hutch soon after 
the optical handling described in Subsection~\ref{ss-inj}. An entrance window is made of 4 mm-thick UV grade synthetic 
fused silica, which is not AR coated but still transmits more than 90\% of a UV or deep UV light. The window is 
mounted to an ICF-114 flange, ensuring an effective diameter of 60 mm, which is large enough compared with a limit 
from the beamline aperture. Air leakage at a joint between the silica window and the SUS flange is severely minimized 
for the use with an ultrahigh vacuum. The entrance window port is located on a vacuum chamber, where an aluminum 
coated silicon mirror with a diameter of 100 mm, called the second mirror, is fixed for a horizontal reflection 
with an incidence angle of 45 degree. This mirror is introduced for the purpose to avoid an overlap of the laser 
injection system with a path of the laser-backscattering photon beam. A thickness of the second mirror, whose surface 
precision of $\sim$$\lambda$/2, is set to 19 mm to get rid of a surface distortion by a holding stress. A laser beam 
from the second mirror chamber is further reflected at another aluminum coated silicon mirror, called the first 
mirror, in order to change the beam direction by 90 degree toward the storage ring. While a diameter of the first 
mirror is the same as that of the second mirror, its thickness is reduced to 6 mm because the backscattering photon 
beam also transmits the first mirror producing e$^+$e$^-$ conversions. Because of high thermal load by x-rays from 
a synchrotron radiation, the first mirror is attached to a copper plate holder, which is cooled by a water circulation. 
A contact between the first mirror and the holder plate is made via Gallium-Indium alloy. The holder plate has 
a rectangular hole of 27 mm $\times$ 42 mm to pass the backscattering photon beam, and is mounted inside a dedicated 
vacuum chamber with an angle adjuster. A surface distortion of the first mirror by the attachment to the holder is 
problematic due to its thin thickness, so that a removal of the distortion is checked by monitoring a laser beam 
focus every time when the first mirror is replaced.

\begin{figure}[t]
 \centering
 \includegraphics[scale=0.6]{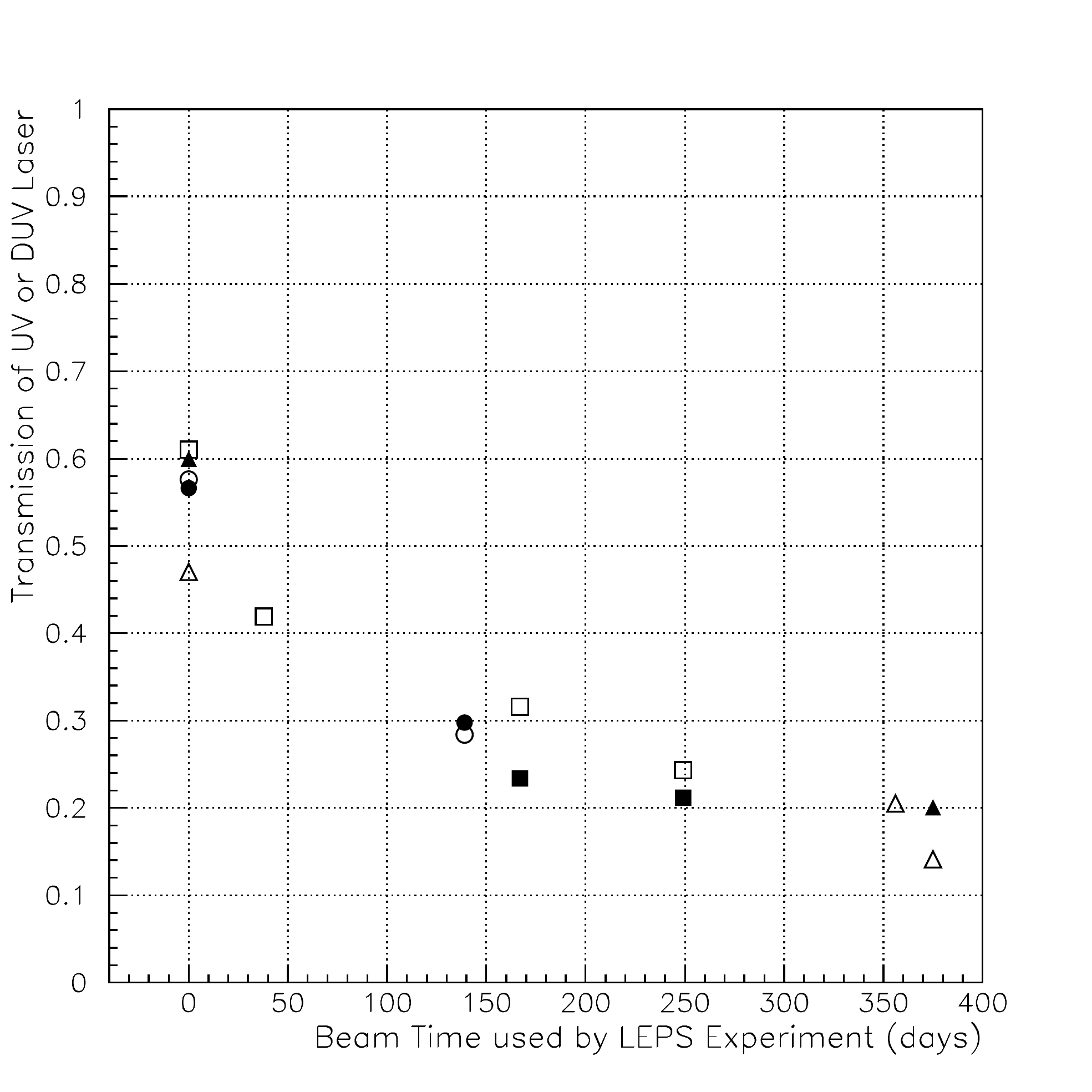}
 \caption{Total transmission rates from the entrance window to the exit of mirror monitor chamber. The transmission 
          rates are plotted as a function of the beam time (in days) used by the LEPS experiment. The day when the 
          entrance window and the first and second mirrors are replaced to new ones is defined to zero in the horizontal 
          axis. The closed (open) symbols correspond to the transmission rates with the UV (deep UV) laser, and the 
          different types of symbol indicate different series of transmission measurements. The absolute values of 
          transmission rates are affected by the common reduction at the monitor mirror and view port to extract 
          a laser beam.}
 \label{fig:trans}
\end{figure}

  The first and second mirrors and the entrance window are damaged under a high radiation environment by x-rays,
which are partly scattered at mirrors. A typical x-ray exposure at the first mirror reaches $\sim$10$^{16}$ 
photons/(sec$\cdot$mm$^2$$\cdot$mrad$^2$$\cdot$0.1\%bw) in the energy region of 1$\sim$100 keV. Attachments of 
hydrocarbons to the mirrors and window from a residual gas under such an environment cause a lower transmission 
rate of laser beam propagation and a distortion of laser beam focus. Therefore, ultrahigh vacuum of $\sim$10$^{-8}$ 
Pa is kept around those optical components by preparing an ion pump under the first mirror chamber and NEG pumps 
on the tops of the two mirror chambers. Nevertheless, it is difficult to stop the radiation damage as seen in 
Fig.~\ref{fig:trans}, which shows relative changes of a total transmission rate of a laser beam through the entrance 
window, the two mirrors, and a laser extraction mirror port. In the figure, the transmission rates were calculated 
by taking a ratio of injected and extracted laser powers, measured by a thermal sensor power meter. The individual 
measurements were done after an arbitrary length of experimental period from a new installation of the optical components. 
Generally, the transmission rate drops to half (one third) in 150 (350) days of radiation exposure. In order to keep 
a reasonably high intensity of the photon beam, the first and second mirrors and the entrance window are replaced 
to new ones each 1$\sim$2 years. After the replacement, the vacuum chambers and ducts, made of SUS, are baked at 
100$\sim$200 $^\circ$C for a whole day to eliminate outgas sufficiently and make a ultrahigh vacuum.

  The injected laser beam must be transported to the straight section, which is located 37 m downstream from the 
laser injection system, and collided with an electron beam with average sizes of $\sigma_x$$\sim$295 $\mu$m and 
$\sigma_y$$\sim$12 $\mu$m. In order to transport the laser beam successfully, a beam position is first checked at 
a monitor mirror port, which is located 1.2 m downstream of the first mirror inside the laser hutch. A BK7 mirror 
is inserted to the laser path in the vacuum by using a handle, and the laser beam is extracted from a view port. 
By monitoring the extracted laser spot relative to a reference mark on a screen attached to the laser hutch wall, 
the rotation and elevation angles of the third and fourth mirrors are adjusted to obtain a correct beam axis. 
Then, this monitor mirror is raised up from the laser path, and another screen monitor, which is located 15.2 m 
downstream of the first mirror inside the storage ring tunnel, is alternatively lowered in the vacuum by a remote 
control. This screen is made of a Desmarquest plate and monitored by a CCD camera through a view port. The angles 
of the third and fourth mirrors are readjusted by using the stepping motors. After this adjustment, the laser beam 
is further transported to a beam end, which is located at the downstream of the far-side bending magnet but on 
the extension line of the straight section. At the beam end, a laser beam is extracted from a vacuum chamber 
through a view port, and a laser spot is monitored on a fluorescent screen using a CCD camera in order to perform 
a close-to-final position adjustment by the mirror angles. Finally, the best position of laser beam axis is 
determined by maximizing a collision rate of laser photons and electrons.

\begin{figure}[t]
 \begin{minipage}{0.5\hsize}
  \centering
  \includegraphics[scale=0.37]{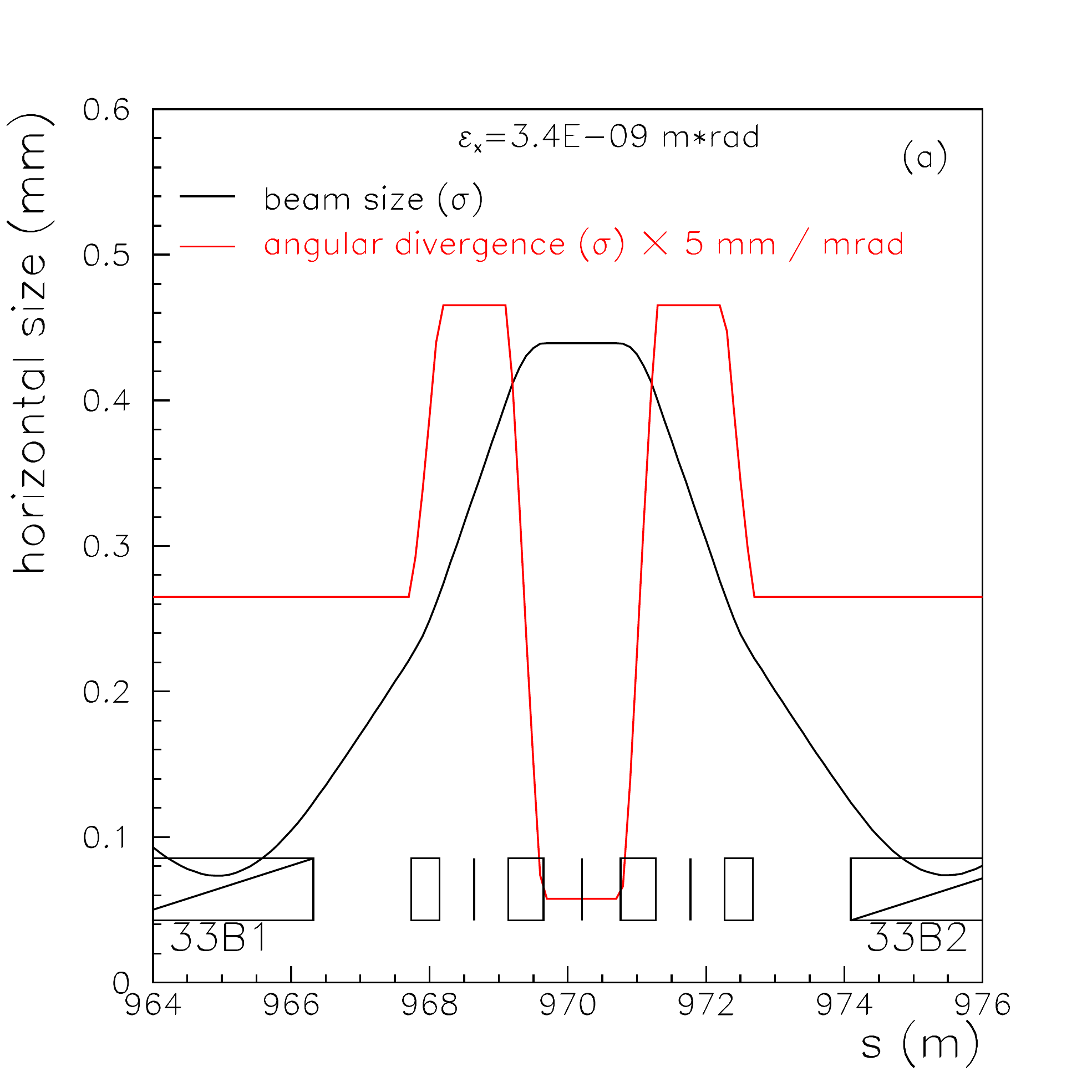}
 \end{minipage}
 \begin{minipage}{0.5\hsize}
  \centering
  \includegraphics[scale=0.37]{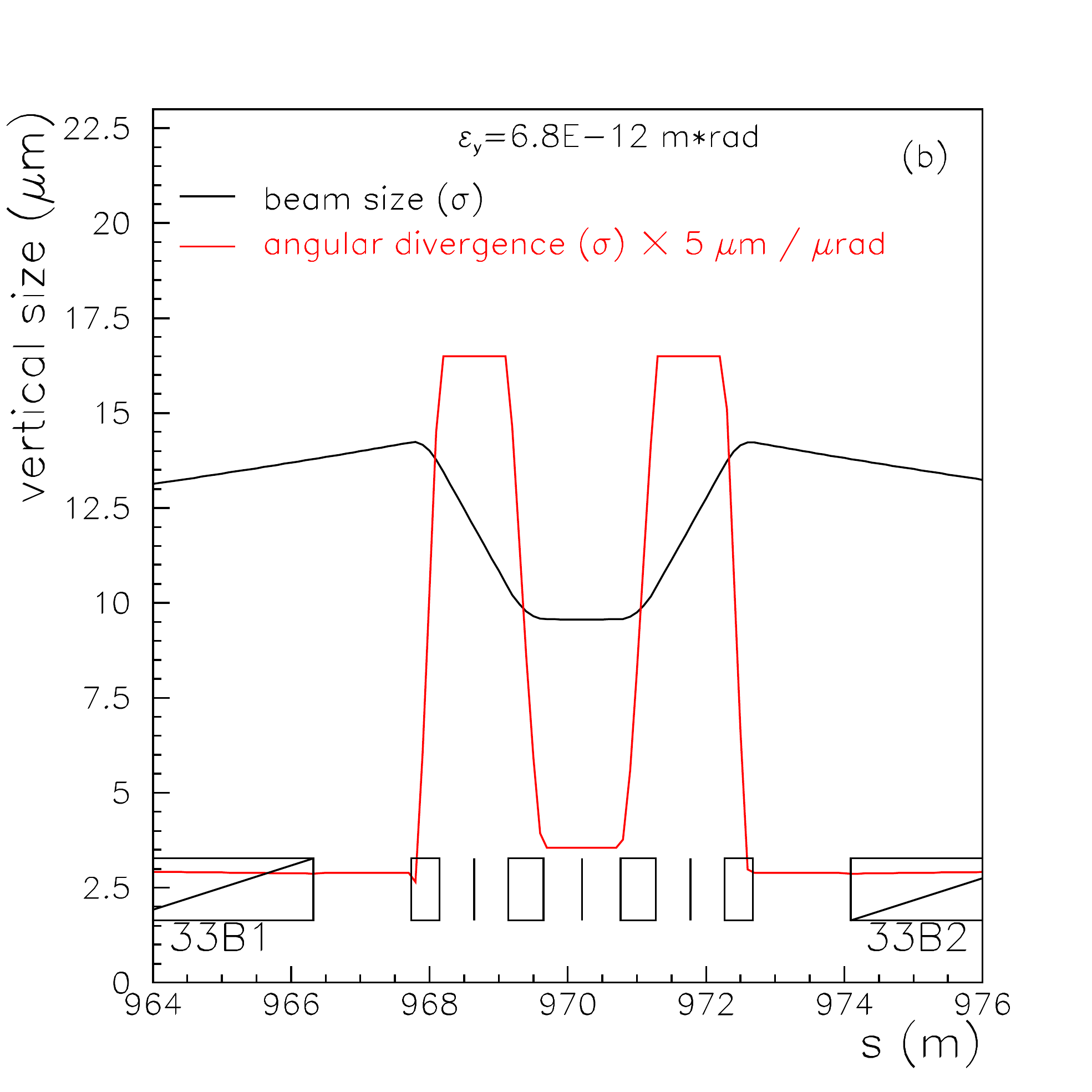}
 \end{minipage}
 \caption{Sizes (black lines) and angular divergences (red lines) of an electron beam at the 7.8 m-long straight 
          section of SPring-8. They are plotted as a function of an arbitrary distance along the storage ring. The
          horizontal and vertical profiles of the electron beam are separately displayed in (a) and (b). Both of
          the size and divergence are shown in $\sigma$. A value of the angular divergence can be recognized in
          mrad for (a) and $\mu$rad for (b) by dividing the vertical scale by 5. The squares in the lower part 
          indicate locations of magnets. }
 \label{fig:ebmdiv}
\end{figure}

  The laser beam path is limited near the storage ring because the vacuum beamline must pass 
bending, quadrupole, and sextupole magnets for the electron beam. There are also an x-ray mask and a crotch 
absorber to eliminate a peripheral part of synchrotron radiation. While the beamline apertures are large enough 
for one laser injection, only 1.4$\sigma$ of the horizontal beam profile can reach the straight section in the 
case of two UV laser injection. At the 7.8 m-long straight section between the two bending magnets, the $\sigma$'s 
of an electron beam size and an electron angular divergence behave as shown in Fig.~\ref{fig:ebmdiv} (a) and (b) 
in horizontal and vertical directions, respectively, with the influence of beam focusing magnets. At the center of 
the straight section, the horizontal and vertical size of the electron beam are increased to 440 $\mu$m and reduced 
to 10 $\mu$m, respectively, while the angular divergences drops to a local minimum in both directions. Here the 
electrons backwardly scatter the laser photons, whose beam size is shrank to focus on the electron beam. 
As described in Subsection~\ref{ss-inj}, the focus point of the laser beam is scanned along 
the straight section by adjusting the micrometer of beam expander to search for the maximum collision rate.
An average divergence of the electron beam at the straight section is 58 $\mu$rad in the horizontal direction,
and this determines a position ambiguity of a recoil electron in proportion to a distance from the collision
point to the tagging detector. In the case of the LEPS experiment, an energy resolution of a backscattering 
photon is dominated by the electron beam divergence, as discussed in the next subsection. This situation is
common in the laser-backscattering beam facilities.

  A laser polarization is measured at the beam end by inserting a stage, where a detector system of a polarizer 
and a photodiode is set up, onto the laser beam axis through a remote control of a pulse motor. A Glan-Laser 
prism, which is usable for high power UV and deep UV lasers, is used as the polarizer. It transmits a linearly 
polarized light by picking out only a vector component parallel to a specific direction which a birefringent 
crystal possesses. The polarizer is rotated by two rounds using a stepping motor remotely, and its light output 
is monitored at each 1 degree by the Si photodiode with a pre-amplifier, whose signal is recorded by ADC.
The ADC values as a function of the rotation angle is fitted by a sine curve with correction factors for detector
responses, and the polarization degree (P$_{laser}$) is defined by:
\begin{equation}
    P_{laser} = \frac{maximum \; ADC \; value - minimum \; ADC \; value}{maximum \; ADC \; value + minimum \; ADC \; value}.
\end{equation}
A direction of linear polarization, which has been aligned vertically or horizontally using a half-wave plate,
is obtained from a phase of the fitted sine curve. The laser polarization measured at the beam end is close to
100\%, and it is taken over to a backscattering photon beam at a collision with an electron following Eqn.~(16)
of the literature by D'Angelo et al. \cite{bcscomp} Here, most of the polarization is transferred near the Compton 
edge, while the polarization degree of the scattering photon decreases as the photon energy becomes low. The photon 
beam polarization is finally calculated from this equation and the measured laser polarization.

\subsection{Properties of the Backscattering Photon Beam} \label{ss-prop}

\begin{figure}[t]
 \begin{minipage}{0.5\hsize}
  \centering
  \includegraphics[scale=0.38]{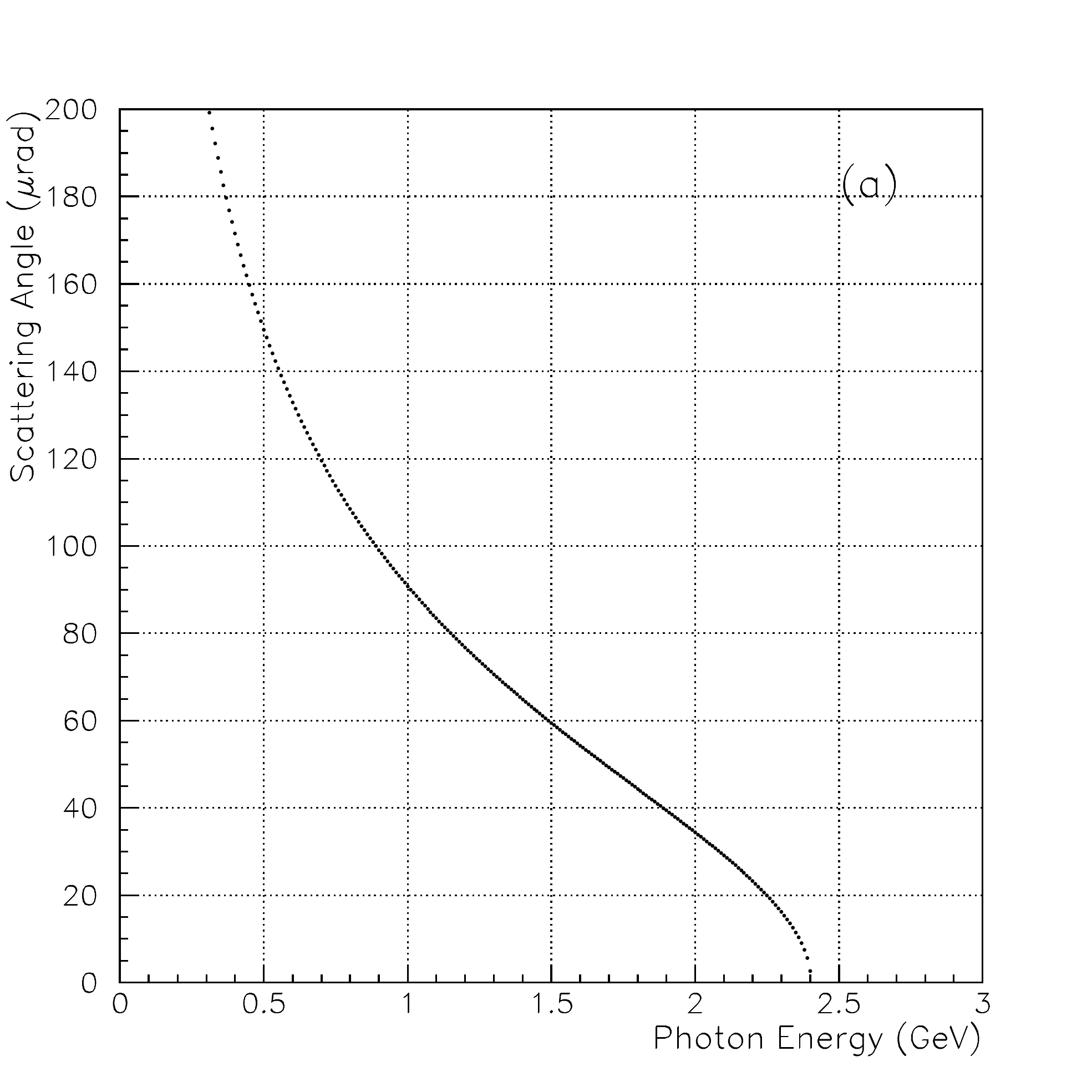}
 \end{minipage}
 \begin{minipage}{0.5\hsize}
  \centering
  \includegraphics[scale=0.38]{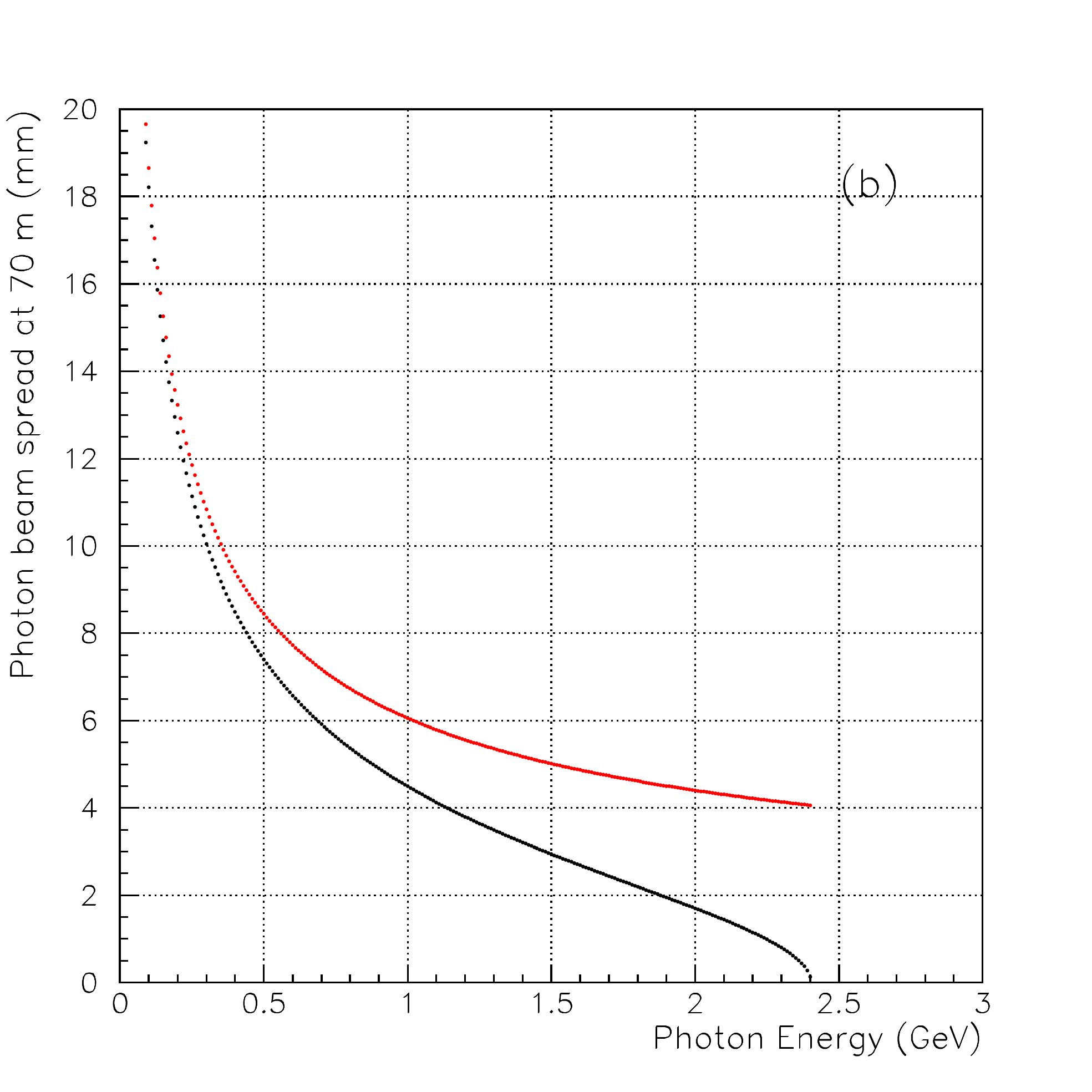}
 \end{minipage}
 \caption{(a) Polar angle of a scattering photon as a function of its energy in the case of UV laser injection
          at the LEPS experiment. This scattering angle is calculated purely from the kinematics of backward 
          Compton scattering. (b) Photon beam size at the experimental hutch as a function of a photon energy.
          A photon distribution, which is calculated on the XY plane at the 70 m downstream using (a), is 
          projected to the X axis, and a RMS of the projection is defined as the beam size. Black dots represent 
          a result of such a raw calculation, while an average angular divergence of the electron beam in the 
          horizontal direction is smeared in red dots. A vertical size of the photon beam must be close to the 
          black dots because of a small angular divergence.}
 \label{fig:bcsang}
\end{figure}

  Photon energies at the backward Compton scattering kinematically depend on a polar angle of scattering 
direction, as shown in Fig.~\ref{fig:bcsang} (a). However, the scattering photon angles are small enough to 
produce a collimated beam at the experimental hutch as shown in Fig.~\ref{fig:bcsang} (b), where a root mean 
square of horizontal projections of the scattering photons is drawn by black dots as a function of the photon 
energy. In reality, this beam spread is smeared by the electron beam divergence of $<\sigma_{x'}>$$=$58 $\mu$rad 
as shown by red dots in Fig.~\ref{fig:bcsang} (b), resulting in a horizontal beam size of $\sigma$$\sim$4.5 mm
in the energy range of 1.5$\sim$2.4 GeV. In contrast, a vertical beam size is reduced to $\sigma$$\sim$2 mm 
because of better electron beam divergence in the vertical direction ($<\sigma_{y'}>$$=$1.5 $\mu$rad).

\begin{figure}[t]
 \centering
 \includegraphics[scale=0.45]{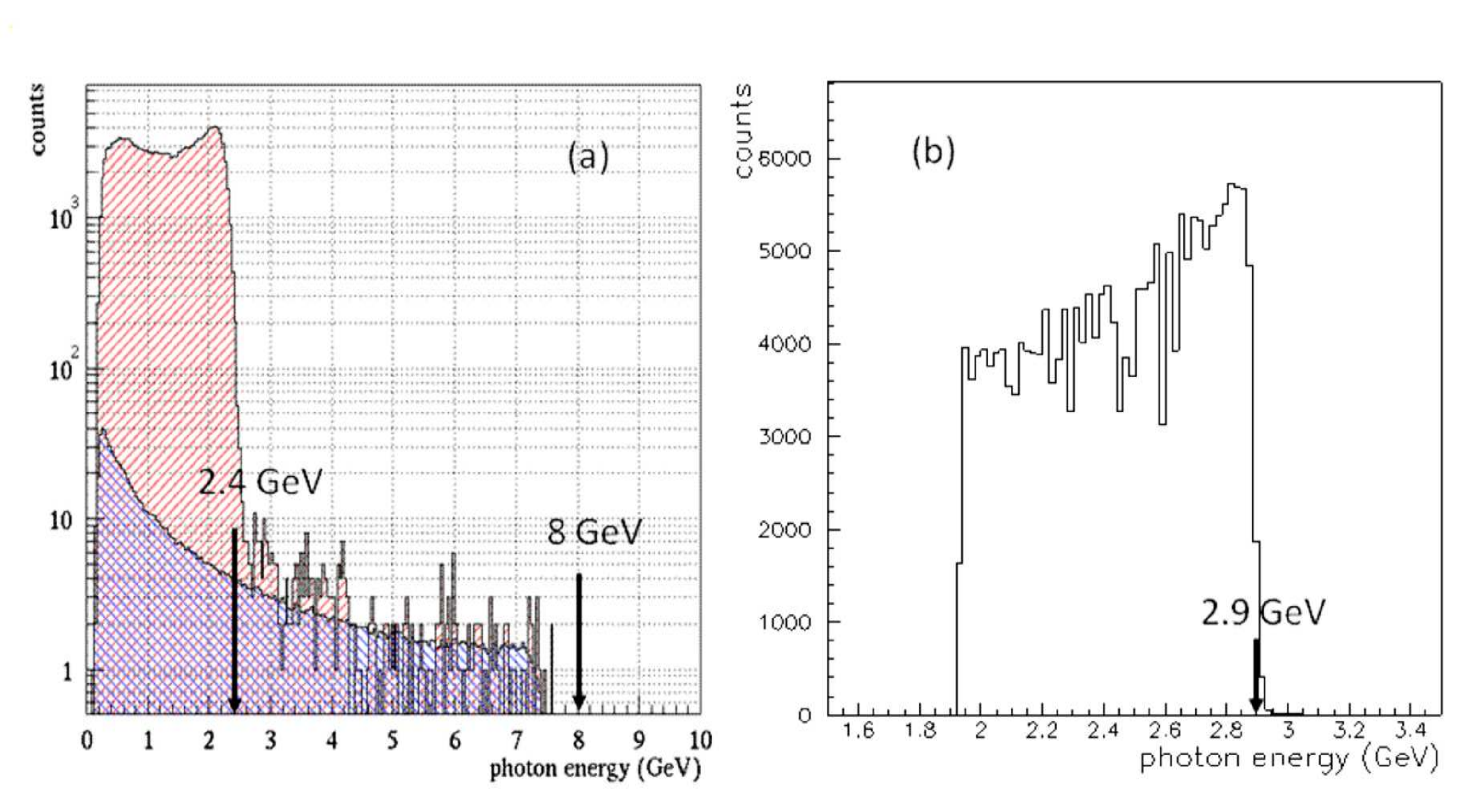}
 \caption{Photon energy spectra for the injections of (a) multiline UV Argon laser and (b) 266 nm-wavelength 
          deep UV laser. A spectrum for Bremsstrahlung photons is overlaid in (a). The spectrum in (b) is largely
          fluctuated because of variations in SSD strip efficiencies. The energy region lower than 1.9 GeV is
          not plotted in (b) because of a SSD module problem at this measurement.}
 \label{fig:espect}
\end{figure}

  Figure~\ref{fig:espect} (a) and (b) show energy spectra of the backscattering photons in the cases of injecting 
a multiline UV Argon laser and a 266 nm-wavelength deep UV laser, respectively. The photon energies in 
Fig.~\ref{fig:espect} (a) are measured by a calorimeter made of PWO crystals, which are specially installed at 
the experimental hutch. A spectrum of a Bremsstrahlung photon beam without a laser injection is overlaid 
by normalizing the two spectra at the highest energy region. It is shown that an intensity of the backscattering 
photon beam is more than two orders larger than that of the Bremsstrahlung beam. On the other hand, the photon 
energies in Fig.~\ref{fig:espect} (b) are measured at a tagging detector to analyze recoil electron momenta, 
called a tagger. A relation between the recoil electron momentum and the photon energy is calibrated by detecting 
pair creations at the forward spectrometer of the LEPS experiment and by comparing a sum of e$^+$ and e$^-$ momenta 
with a hit position information at the tagger. As expected, the Compton edges of the UV and deep UV laser injections 
correspond to 2.4 GeV and 2.9 GeV, respectively, and both the energy spectra are continuously distributed without 
a change of order.

  During usual experiments, the photon energies are measured event by event at the tagger with a coincidence of
a trigger signal from the LEPS spectrometer. At early experiments, the tagger was composed from 10 fingers of 
trigger scintillators and two layers of 100 $\mu$m-pitch SSD strips to analyze a recoil electron path, which 
deviates from the electron beam trajectory by the 0.68 Tesla bending magnet depending on a recoil momentum. 
Basically, a 1 mm deviation at the tagger corresponds to a 30 MeV/c difference of the recoil electron momentum. 
The photon energy is evaluated by subtracting the recoil momentum from 7.975 GeV under the energy conservation 
at the backward Compton scattering. Because of the acceptance of the tagger, a recoil electron can be detected 
with a momentum less than 6.5 GeV/c, so that the backscattering photons are tagged at E$_{\gamma}$$>$1.5 GeV. 
A resolution of the measured photon energy is not dominated by the position resolution of the tagger, whereas 
it mostly comes from the electron beam divergence. In a simple estimation, a horizontal divergence of 58 $\mu$rad 
causes a position uncertainty of 0.4 mm at the tagger, which is about 7 m downstream of a collision point, resulting 
in a photon energy resolution of 0.4 mm $\times$ 30 MeV/mm $=$ 12 MeV. This is consistent with an estimation from 
real data, where the LEPS spectrometer detects two charged tracks of the K$^+$K$^-$p or $\pi^+\pi^-$p final state 
in $\gamma$p reactions. Here, a photon energy is backwardly deduced by assuming a mass of the missing particle, 
and compared with the tagger energy by taking into account a contribution from spectrometer resolutions, estimated 
in a MC simulation. From the year 2005, the SSD part of the tagger has been 
replaced to 1 mm-square scintillation fibers, which are placed with a 0.5 mm shift at different layers. A position 
resolution of the new tagger is expected to be less than $0.5/\sqrt{12}$ mm with a pattern recognition, and the 
electron beam divergence still dominates the photon energy resolution.

  An intensity of the laser-backscattering photon beam is proportional to a laser photon flux, an electron beam
current, and a Compton scattering cross section. The laser photon flux is derived by dividing a laser output 
power by a laser photon energy, which is calculated from $Planck \; constant \times light \; speed \; / \; laser 
\; wavelength$. Therefore, a ratio of photon fluxes from the 355 nm UV laser and the 266 nm deep UV laser becomes 
(8 W / 1 W) $\times$ (355 nm / 266 nm) $\sim$ 11. In addition, the differential cross section of backward Compton 
scattering for the wavelength of 355 nm is $\sim$0.20 barn/GeV, which is larger than $\sim$0.15 barn/GeV for 266 
nm. In contrast, a tagged ratio in the total photon intensity for the 266 nm laser is 47\% based on a fraction of 
the Compton scattering spectrum above 1.5 GeV, while that for the 355 nm laser is reduced to 38\%. In total, the 
tagged photon beam intensity with the 266 nm laser is expected to be about 9\% of that with the 355 nm laser. As 
already described in Section~\ref{ss-inj}, the measured tagged intensities were typically $\sim$1.5$\times$10$^6$ 
sec$^{-1}$ and $\sim$1.5$\times$10$^5$ sec$^{-1}$ for the injection of a mode-lock UV laser and a solid-state 
deep UV laser, respectively. The observed intensity ratio was consistent with the above estimation based on the
theoretical formulation of a photon energy spectrum. A simultaneous injection of two lasers is now available, and 
the tagged photon intensities are raised up to 2$\sim$3$\times$10$^6$ sec$^{-1}$ and 2$\sim$3$\times$10$^5$ 
for the UV and deep UV lasers, respectively. The magnification factors of $\sim$1.7 have been observed with
acceptance corrections for the beamline aperture. The slight reduction from a factor of two is caused by 
a deviation from a true head-on collision with a small injection angle. 

  The electron beam in the storage ring possesses a life time depending on the filling pattern of electron bunches. 
Generally, an isotropic filling with a few hundreds bunches results in the electron beam life more than 200 hours, 
while existences of single bunches shorten the life to a few tens hours. If there is no further injection of electrons, 
the intensity of a backscattering photon beam drops following the decrease of the electron beam current. In addition, 
the collisions of photons and electrons raise up a decreasing speed by loosing electrons with a rate of the photon 
beam intensity. The reduction of electron beam life varies typically in the range of 5$\sim$30\%, and a longer life
time of the electron beam is more influenced by the backward Compton scattering. At early operations of SPring-8, 
an average current of the electron beam was therefore lower than 100 mA, and electrons were refilled each half day. 
However, from the year 2004, Spring-8 has started a top-up operation, which continuously fills electrons every one 
minute. Currently, an electron beam with 100 mA is stably provided about 5,000 hours per year, and the top-up
operation contributes to a net gain of the backscattering photon beam intensity. This operation has also solved 
the life reduction problem due to the backward Compton scattering beamline, so that other x-ray beamlines under 
co-operations have received no more influences.

  The photon beam travels about 70 m backwardly along the injected laser path from the straight section to the 
experimental hutch. The beamline with a ultrahigh vacuum lasts until the laser hutch, and ends at a 0.55 mm-thick 
aluminum window. At this exit window, lead plates with a total thickness of 1.5 mm are additionally mounted as 
an x-ray absorber. In contrast with the x-ray reduction at the absorber, an additional contamination in the photon 
beam arises from e$^+$e$^-$ pair creations, which mainly occur at the first mirror, the aluminum window, and the 
x-ray absorber. Therefore, a sweep magnet, which gives a dipole field of 0.6 Tesla using Neodymium magnets, is placed 
together with a $\phi$25 mm-hole lead collimator 4 m downstream of the aluminum exit window inside the laser hutch. 
Soon after the sweep magnet, a medium vacuum beam pipe ($\sim$1 Pa) with Kapton windows at both ends continues from 
the laser hutch to the experimental hutch. A transmission rate of the backscattering photon beam, which is partly 
lost by the pair creations at the beamline materials from the interaction point to the experimental hutch, is 
estimated to be about 50\% in total.

\subsection{Experimental Setup} \label{ss-det}

  The photon beam delivered to the experimental hutch first passes a lead collimator with a hole of 4 cm 
$\times$ 5 cm. Then, contaminations of e$^+$e$^-$ conversions, which arise at the medium vacuum beam pipe, 
are vetoed at a 5 mm-thick plastic scintillation counter (upveto counter). Finally, the photon beam reaches
a target, where hadrons are photoproduced. The target material is chosen depending on an experimental program 
from either of liquid hydrogen, liquid deuterium, or a solid nuclear target (ex. copper, carbon, polyethylene, 
etc). The liquid materials are filled inside a 15 cm-long cell with Kapton sheet windows. This cell is connected 
to large volume gas cylinders, and only the cell part is cooled down to 18$\sim$20 K inside a vacuum container.

\begin{figure}[b]
 \centering
 \includegraphics[scale=0.55]{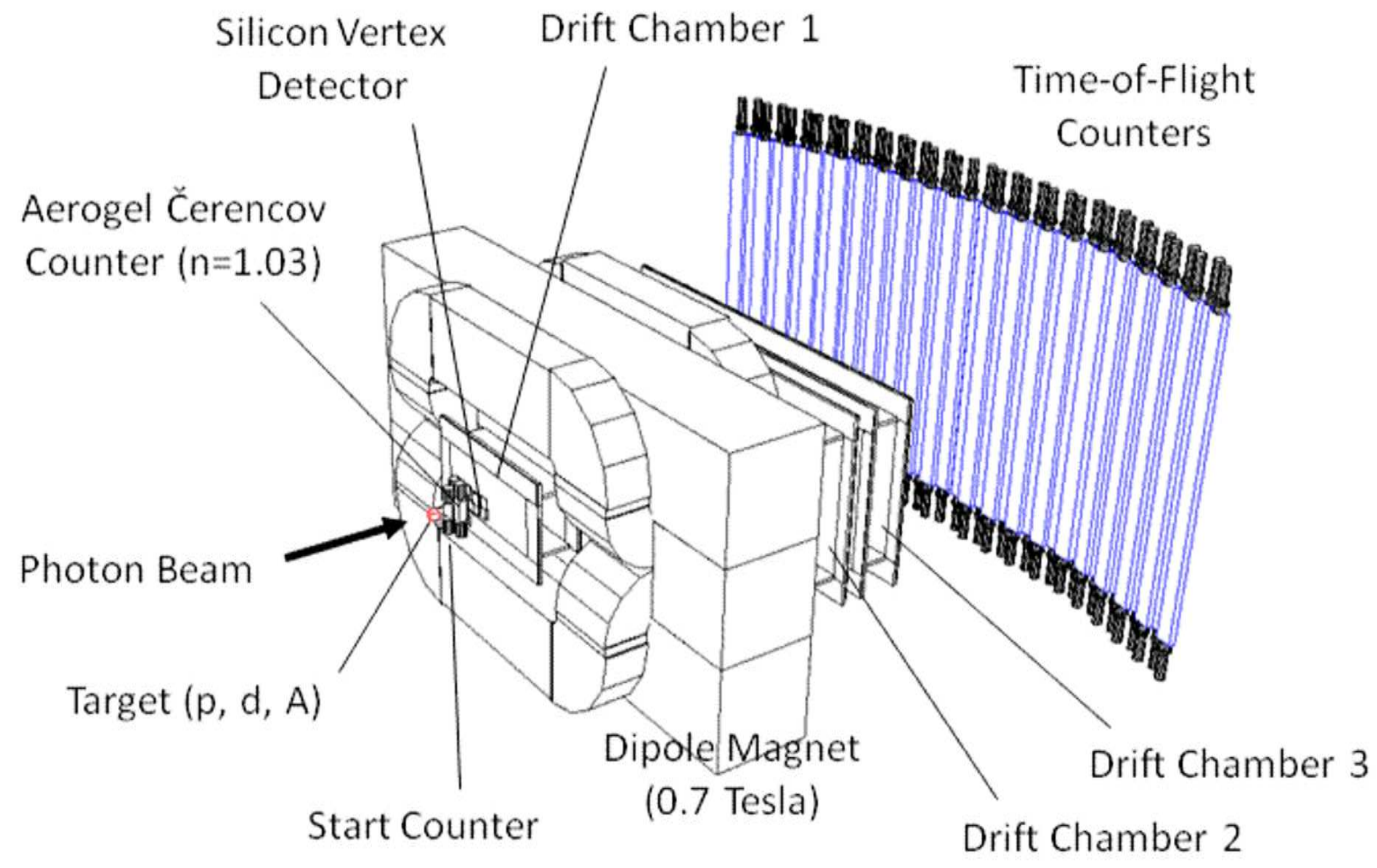}
 \caption{The forward spectrometer setup used at the LEPS experiment. A photon beam comes from the left-hand
          side. The target material is switched to either of liquid hydrogen (p), liquid deuterium (d), or 
          nuclear matter (A).}
 \label{fig:fwdspect}
\end{figure}

\begin{figure}[t]
 \centering
 \includegraphics[scale=0.55]{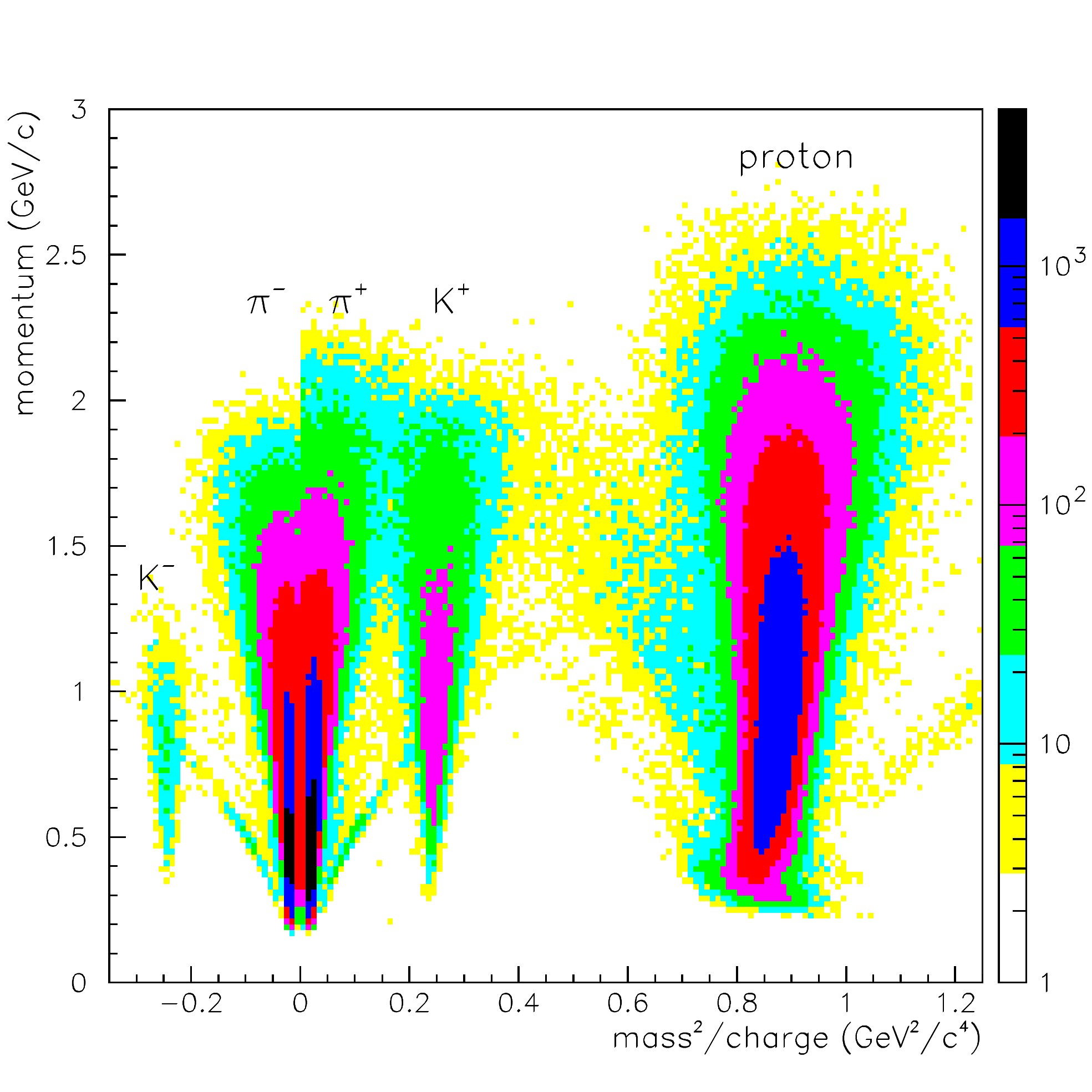}
 \caption{Momentum vs. mass-squared for charged particles detected at the LEPS forward spectrometer.
          Pair creations of e$^+$e$^-$ are reduced by removing the tracks which are close to the horizontal 
          plane including the photon beam axis. The vertical bands correspond to proton, K$^+$, K$^-$, $\pi^+$, 
          and $\pi^-$, respectively. The tilting bands which leave from the vertical bands more at the higher 
          momentum are caused by an improper selection of the RF signal, which determines a start timing of 
          the time-of-flight measurement in each 2 nsec. Contaminations to a selected particle band from others 
          are mostly eliminated after identifying all particles in the final state with detected charged particles
          and their missing mass.}
 \label{fig:pid-mass2}
\end{figure}

  Charged particles produced at the target are detected at a forward spectrometer (LEPS spectrometer), which 
covers geometrical acceptance of $\pm$20 degree and $\pm$10 degree in the horizontal and vertical directions, 
respectively. A setup of the LEPS spectrometer is shown in Fig.~\ref{fig:fwdspect}. A momentum of the charged 
particle is analyzed at a silicon strip vertex detector (SVTX) and three of large frame drift chambers (DC1-3) 
under a dipole magnetic field, which is 0.7 Tesla at the center pole. A momentum resolution is typically 
$\sim$0.6\% for a 1 GeV/c particle. Since e$^+$e$^-$ conversions are caused at the target with a high rate 
compared with that of hadron productions, a threshold-type aerogel ${\rm \check{C}}$erenkov counter, whose 
refractive index is 1.03, is set up between the target and the SVTX for e$^+$e$^-$ veto (AC). The e$^+$ and 
e$^-$ tracks are mostly concentrated along the photon beam direction, and the SVTX is therefore designed to 
have a 5 mm-squared hole to avoid a high current damage. The dipole magnetic field bends the e$^+$ and e$^-$ 
tracks closely on the horizontal plane including the photon beam axis, so that the e$^+$e$^-$ tracks are 
blocked at two iron bars, which are placed on this horizontal plane with a height of 5 cm inside the magnet
(e$^+$e$^-$ blocker). These two bars are separated around the photon beam axis by 15 cm in order to remove
materials on the beam path. Particle types of the charged hadrons are identified from a time-of-flight 
measurement between the target and a wall of 40 plastic scintillation counters (TOF counters). 2 m-long 
counters are vertically aligned 4 m downstream of the target, and a 4 cm-wide vertical slit is prepared at 
a center of the wall to escape the photon beam. A start timing of the time-of-flight measurement is determined 
by a RF signal from the storage ring, where electron bunches are stored with a RF frequency of 508.58 MHz.
A proper RF signal or electron bunch is selected if its timing is closest to a reference time from a 5 mm-thick 
plastic scintillation counter, located right downstream of the target (start counter). Based on the measurements
of momentum (P) and time-of-flight (T), the squared mass (M$^2$) of a detected charged particle is calculated 
as follows:
\begin{equation}
   M^2 = P^2 \times ((T/L)^2-1),
\end{equation}
where L represents a flight length from the target to a TOF counter, obtained by a tracking code. 
A distribution of momentum versus mass-squared is shown in Fig.~\ref{fig:pid-mass2}, and a 3$\sim$4$\sigma$
selection of the mass band is usually applied depending on the momentum to choose desired particle types. 
A mass resolution for a 1 GeV/c kaon is typically 30 MeV/c$^2$, which corresponds to a time-of-flight 
resolution of 150 psec.

  A trigger signal is made from a coincidence of the tagger, the start counter, and either of the TOF counters
with veto signals from the upveto counter and the AC. A trigger timing is determined by the start counter, and
a trigger signal from the TOF counters is widened enough for slow particles. For a tagged photon intensity of 
$\sim$10$^6$ sec$^{-1}$, a trigger rate is typically counted to be 100$\sim$200 sec$^{-1}$ including e$^+$e$^-$ 
contaminations. A large dead time due to a high rate counting is caused at the tagger, and 10$\sim$50\% of recoil 
electrons are unaccepted for a tagged intensity of 10$^6$ sec$^{-1}$ depending on the filling pattern of electron 
bunches. An isotropic filling pattern generally reduces the tagger dead time. Sometimes multiple hits are recorded 
at the tagger in a triggered event, arising from a shower at the inner wall structure of the tagger, 
simultaneous $\gamma$e scatterings from one bunch, and so on. This type of events makes it difficult to 
select a true hit, which must determine a recoil electron momentum or a photon energy. Such a situation 
happens in $\sim$20\% of collected events even with a tight timing requirement offline.

  Depending on physics programs, a time projection chamber (TPC) is additionally set up right upstream of the 
forward spectrometer. The TPC possesses a hexagonal cross section with a length of about 0.9 m, and is held inside 
a 2 Tesla solenoidal magnet, whose bore diameter is 600 mm. A target material is placed at an upstream part of 
a TPC inner hole, and the TPC is used as a side-way tracker for charged particles. A combination of the TPC
and the forward spectrometer covers $\sim$90\% of 4$\pi$ solid angles. More details of the TPC are described 
elsewhere \cite{tpcdet}.

  Photons which do not interact with the target materials stop at a beam dump at the most downstream part of 
the experimental hutch. The beam dump is made of a 400 mm-thick lead block and surrounded by concrete walls.
In the experimental hutch, rate monitors, made from plastic scintillators and converters, are also installed 
just upstream of the beam dump. They are prepared as a supplementary means for counting a photon beam intensity.

\subsection{Physics Results} \label{ss-phy}

\begin{figure}[b]
 \centering
 \includegraphics[scale=0.6]{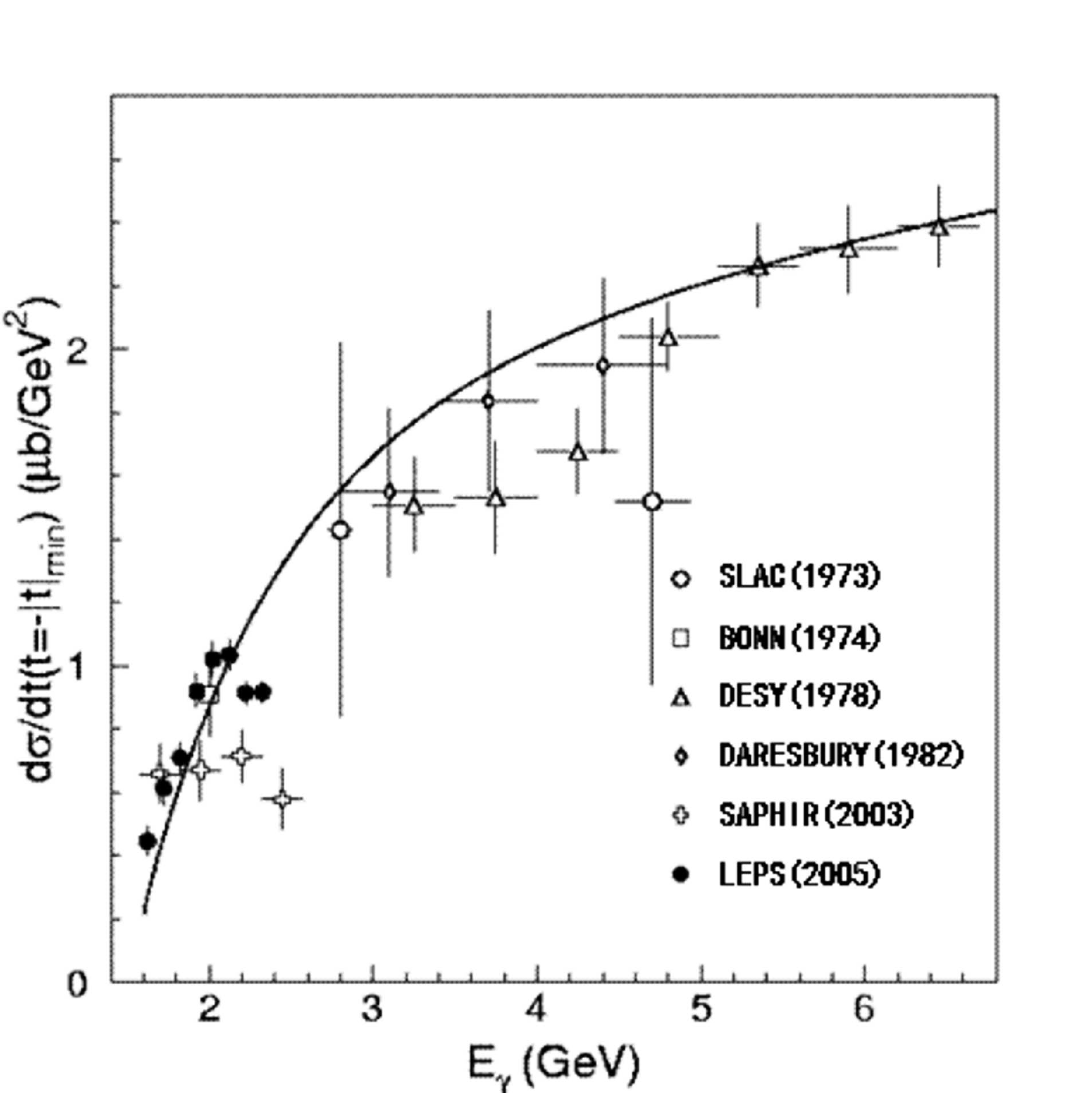}
 \caption{Differential cross section of $\phi$ photoproduction off proton at the minimum momentum transfer t.
          The solid line indicates a theoretical calculation based on a conventional pomeron exchange, $\pi$ 
          and $\eta$ exchanges, N$^*$ resonance contributions, and a s$\bar{\rm s}$ knockout \cite{phitheor}. 
          The closed circles show the result from the LEPS experiment, indicating a local maximum at 
          E$_\gamma$$\sim$2.0 GeV \cite{phip}.}
 \label{fig:phixsec}
\end{figure}

  The LEPS experiment has advantages to investigate t-channel photoproduction of a meson and baryon pair thanks 
to the detector acceptance which covers extremely forward angles and the high polarization degree of the photon 
beam. Spin observables are defined by angular distributions of a meson or its decay products relative to 
a polarization vector of the linearly polarized photon beam, and depend on the nature of an exchanged particle 
in the t-channel. A good example of such observables is seen in photoproduction of a $\phi$ meson, which is 
identified from invariant mass of a K$^+$K$^-$ pair detected at the forward spectrometer. If the $\phi$$\to$
K$^+$K$^-$ decay plane tends to be parallel (perpendicular) to the polarization vector in the azimuthal angle
direction at the $\phi$ rest frame or the Gottfried-Jackson frame, a natural (unnatural) parity exchange process, 
corresponding to mediating pomerons and scalar mesons (pseudoscalar mesons), must be enhanced with a positive 
(negative) spin density matrix element $\rho^{1}_{1\;-1}$. In the $\phi$ photoproduction from a free proton
target, this element was measured to be $\rho^{1}_{1\;-1}$$\sim$0.2, which suggests a natural parity exchange
contribution is large \cite{phip}. The incoherent $\phi$ photoproduction from a nucleon inside a deuteron target 
provided a similar value of $\rho^{1}_{1\;-1}$ \cite{phin}. In contrast, $\rho^{1}_{1\;-1}$$\sim$0.5 was 
observed in the coherent $\phi$ photoproduction from a deuteron, suggesting dominance of natural parity exchange 
processes \cite{phid}. This result is consistent with the fact that an isovector pion cannot couple to an 
isoscalar deuteron target. In all the above $\phi$ photoproductions, a polar angle distribution of K$^+$ at 
the $\phi$ rest frame obeys $\sin^2\theta$, and helicity-conserving processes were found to be dominant near 
the threshold and at the small momentum transfer t. A more systematic measurement of the spin density matrix 
are described in the literature by Chang et al. \cite{phisdm}, observing small but finite helicity-nonconserving 
amplitudes. In addition to the spin observables, differential cross sections were measured for comparisons with 
theoretical calculations including pomeron, $\pi$, and $\eta$ exchanges, N$^*$ resonance contributions, and 
a s$\bar{\rm s}$ knockout process. Since the conventional contributions are relatively reduced near the threshold, 
an exotic contribution may be probed in the LEPS energy range. Figure~\ref{fig:phixsec} shows photon energy 
dependence of the differential cross section at t$_{min}$, and a local maximum at E$_\gamma$$\sim$2.0 GeV was 
observed \cite{phip}. This phenomenon has not been explained yet, requiring to refine theoretical understandings. 
The $\phi$ photoproductions from nuclear targets were also investigated in terms of a medium modification effect, 
and a large $\phi$N cross section of $\sigma_{\phi N}$$\sim$35 mb was observed \cite{phia}.

  Hyperon photoproduction with a forward K$^+$ detection is another example of spin observable measurements.
An azimuthal angle distribution of K$^+$Y production plane relative to the photon polarization vector depends 
on an exchanged particle in the t-channel. This dependence is described by a quantity called `photon beam 
asymmetry' ($\Sigma$), which becomes positive (negative) when the production plane tends to be perpendicular (parallel) 
to the polarization vector. Generally, a $K^*$ (K) exchange provides a positive (negative) photon beam asymmetry, 
while s- and u-channels force the asymmetry to be around zero. Photoproductions of ground-state $\Lambda$ and 
$\Sigma^0$ from a proton target suggest $\Sigma$=0.2$\sim$0.6 depending on a center-of-mass energy and a momentum 
transfer \cite{lamsig-1,lamsig-2}. This indicates a large contribution of $K^*$ exchange and moderate contributions 
from K exchange and s-channel resonances, as recognized also from differential cross sections. Photoproduction of 
$\Sigma^-$, which is a different isospin component for $\Sigma^0$, was also investigated in the reaction $\gamma$n
$\to$K$^+$$\Sigma^-$ from a deuteron target, and the photon beam asymmetry was measured to be roughly twice of 
that in the $\Sigma^0$ photoproduction, suggesting the dominance of K$^*$ exchange contribution \cite{sigm}. 
On the other hand, the K$^*$ contribution was found to be suppressed with a small and negative photon beam 
asymmetry in photoproduction of a hyperon excitation $\Sigma^-$(1385)
off a neutron inside a deuteron target \cite{sig1385}. A small value of photon beam asymmetry was similarly 
measured in photoproduction of $\Lambda$(1520) from a proton target, and the asymmetry was found to grow up
along with an increasing photon energy \cite{lam1520-1,lam1520-2}. A coupling constant among a K$^*$, a nucleon, 
and a hyperon excitation was recognized to be small at low energies. The hyperon photoproductions are interesting 
also to search for missing N$^*$ resonances, which may couple more strongly to strangeness. A few measurements of 
differential cross sections have indicated existences of such resonance contributions \cite{lamsig-2,sig1385}. 
A bump structure was observed as well in the $\Lambda$(1520) photoproduction cross section at E$_\gamma$$\sim$2.0 
GeV and a small momentum transfer \cite{lam1520-2}, while it was not clear in a large momentum transfer region 
\cite{lam1520-1}. Interpretations of this structure are controversial yet.

\begin{figure}[b]
 \centering
 \includegraphics[scale=0.6]{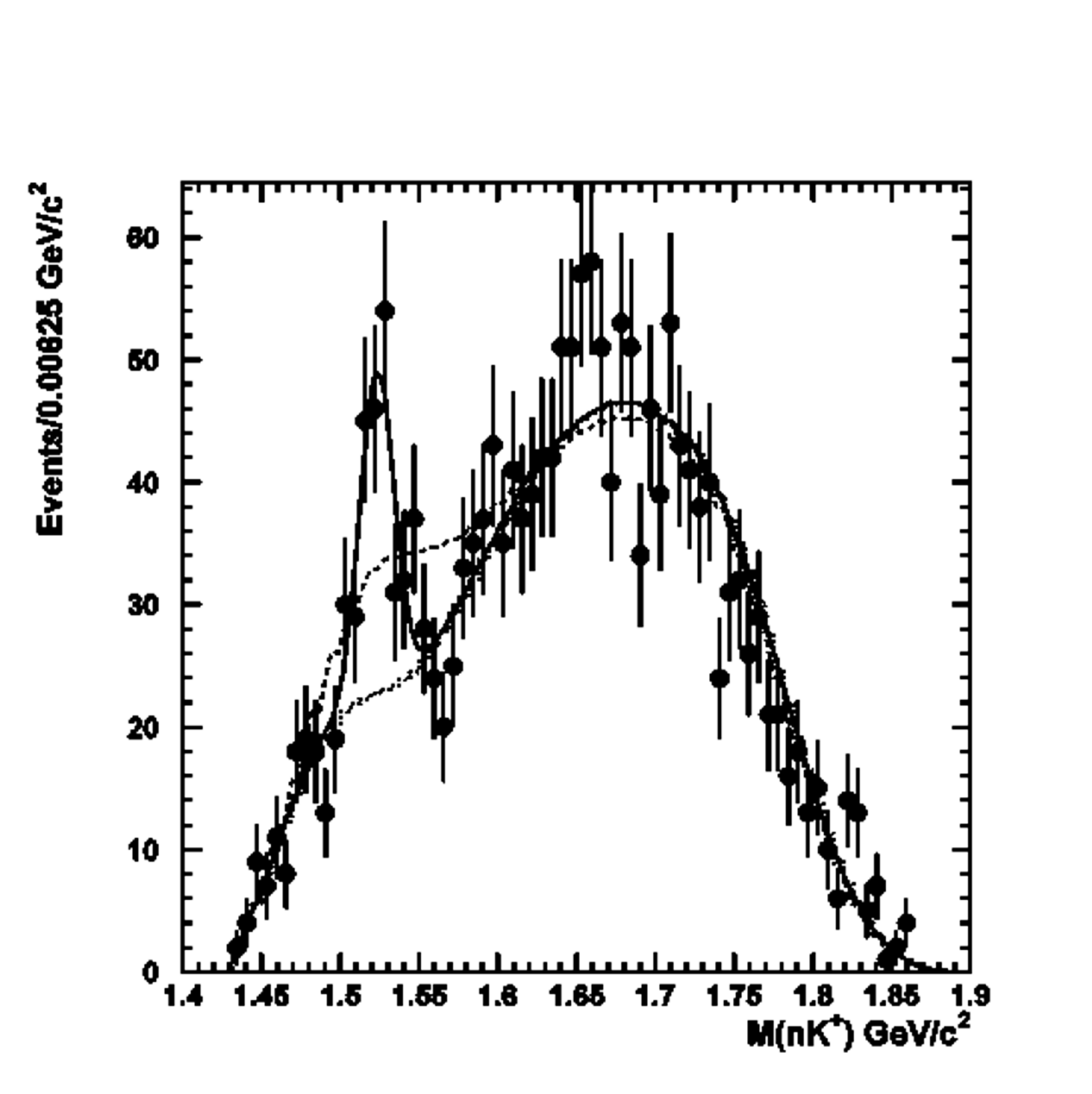}
 \caption{K$^+$n invariant mass distribution in the reaction $\gamma$n$\to$K$^-$K$^+$n off deuteron. A K$^+$K$^-$
          pair was detected at the forward spectrometer, and a neutron momentum vector was solved by assuming
          a spectator proton possessed the minimum momentum in the missing momentum vector, which was carried by
          a pn system \cite{thetad}. By using this method, the effect of Fermi motion is minimized at the K$^+$n
          invariant mass calculation. In the figure, two kinds of fit were performed by assuming both signal and 
          background contributions (solid line) or only a background contribution (dashed line). A statistical 
          significance of 5.1$\sigma$ for the observed peak was obtained from a difference of log likelihoods in 
          the two fits.}
 \label{fig:theta}
\end{figure}

  Geometrical acceptance of the LEPS spectrometer is limited to forward angles, but backward meson photoproductions 
can be simultaneously explored by detecting a forward-going baryon. This type of kinematics is generally useful to 
investigate s- and u-channel processes without a large contribution from the t-channel. For example, backward $\eta$
photoproduction from a proton target can be examined by calculating a missing mass with a forward proton detection 
\cite{eta}. The differential cross sections show a bump structure around 2.1 GeV in the center-of-mass energy, and
it may indicate a N$^*$ resonance which couples to strangeness because the $\eta$ meson possesses a large fraction 
of s$\bar{\rm s}$ content. Similarly, backward photoproduction of $\pi^{0\;}$ \cite{pi0} and forward photoproduction 
of $\Lambda^{\;}$ \cite{lamback} and $\Lambda$(1520)$^{\;}$ \cite{lam1520-1} have been studied to obtain new insights 
of reaction mechanisms. In the $\Lambda$(1520) photoproduction, reactions with both proton and deuteron targets were 
analyzed by identifying the $\Lambda$(1520) in invariant mass distributions of detected K$^-$ and proton. 
The $\Lambda$(1520) production from a neutron was found to be suppressed by comparing differential cross sections from 
the two targets, and it was possibly indicated that a contact term, which coexists with a t-channel K exchange to 
conserve gauge invariance, provides a main contribution to the $\Lambda$(1520) photoproduction making a difference of 
production rates from proton and neutron.

  Probing exotic hadron structures has been another important issue to be analyzed intensively. The LEPS experiment
has first reported an evidence for a pentaquark state $\Theta^+$, which consists of uudd$\bar{\rm s}$ quarks at 
a ground state \cite{thetac}. A peak structure with a narrow width was observed at 1.54 GeV/c$^2$ by detecting a 
K$^+$K$^-$ pair in a reaction $\gamma$n$\to$$\Theta^+$K$^-$$\to$K$^+$nK$^-$ from a carbon target and by correcting 
a Fermi motion effect in a K$^-$ missing mass distribution. In order to confirm the existence of this narrow peak, 
another data set using a liquid deuterium target was further analyzed by constructing a K$^+$n invariant mass 
distribution with reduced influence of Fermi motion, and a corresponding peak structure was observed again with 
a statistical significance of 5.1$\sigma$ as shown in Fig.~\ref{fig:theta} \cite{thetad}. However, the existence 
of $\Theta^+$ is still controversial after worldwide experimental inspections. The LEPS experiment has therefore 
collected three times higher statistics data using high intensity photon beam with a simultaneous two-laser injection, 
and the new data set is being analyzed. Another candidate of exotic hadron is $\Lambda$(1405), which is theoretically 
discussed as a meson-baryon molecular state. The $\Lambda$(1405) photoproduction was investigated using the TPC and 
the LEPS spectrometer in order to detect all the charged products and distinguish the $\Lambda$(1405) from the 
$\Sigma^0$(1385), which is considered to be a conventional three-quark state \cite{lam1405}. A strong energy 
dependence was observed in the production ratio of $\Lambda$(1405) and $\Sigma^0$(1385), possibly suggesting 
exotic internal structure of the $\Lambda$(1405).

\section{New Developments of a GeV Photon Beam} \label{sec-future}

  Techniques to produce a GeV photon beam by backward Compton scattering have been continuously developed at 
the existing facilities. At the LEPS experiment, a maximum photon energy upgrade by introducing a stable deep UV 
laser and a beam intensity upgrade by a simultaneous injection of two UV or deep UV lasers have been achieved as 
described in the previous section. Further developments are as well in progress, and newly available techniques 
are being adopted at a new beamline construction, which aims a next generation hadron experiment, called LEPS2, 
at SPring-8.

\begin{figure}[b]
 \centering
 \includegraphics[scale=0.55]{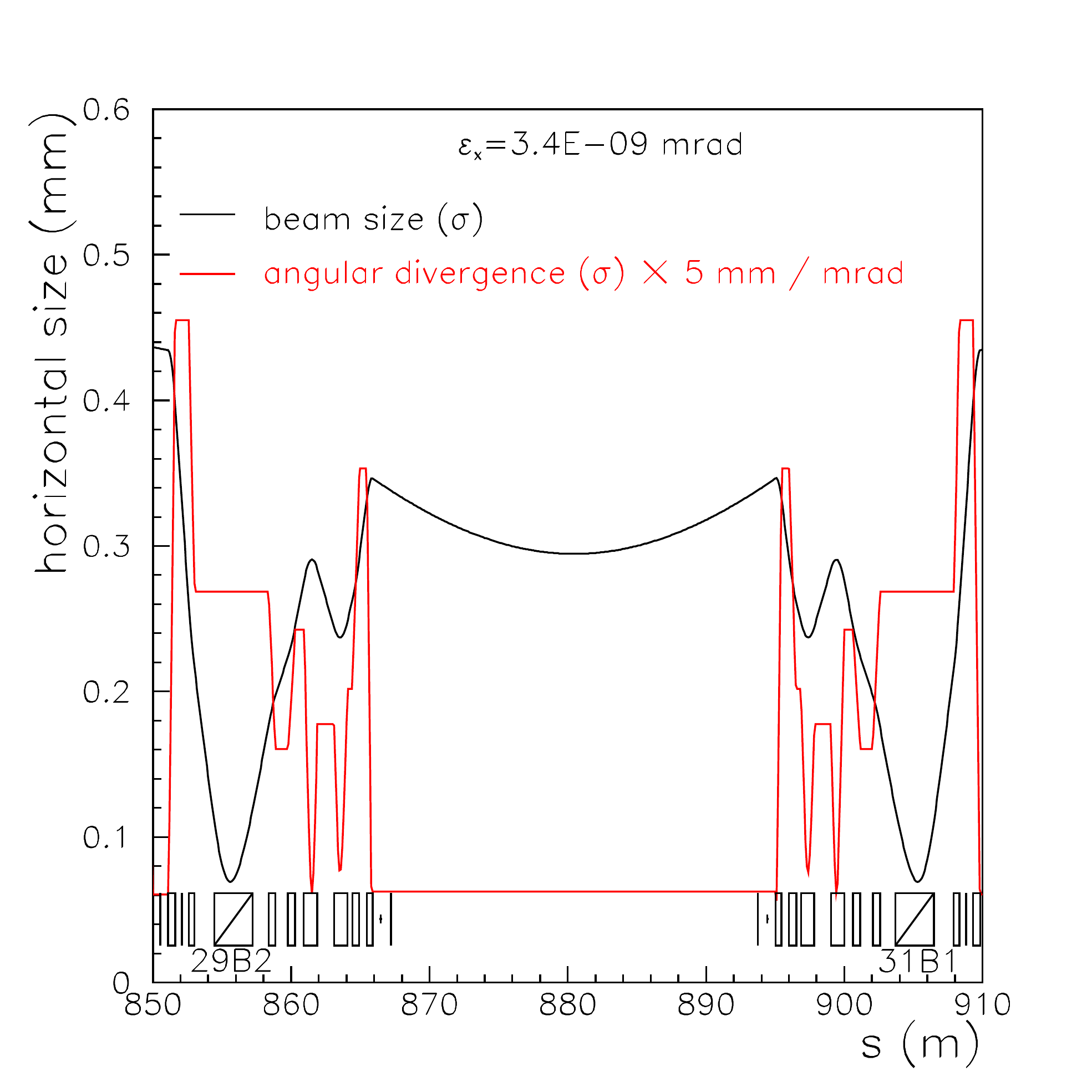}
 \caption{Size (black line) and angular divergence (red line) of the electron beam in the horizontal direction 
          at the 30 m-long straight section of SPring-8 (BL31ID). The horizontal axis represents an arbitrary
          distance along the storage ring. A value of the angular divergence is recognized in mrad by dividing
          the vertical scale by 5. The horizontal divergence is reduced to 12 $\mu$rad, compared with 58 $\mu$rad
          at the 7.8 m-long straight section (BL33LEP).}
 \label{fig:bl31}
\end{figure}

  The LEPS2 project was approved in 2010, and a construction of the facility soon started at the new beamline of
SPring-8 (BL31ID). The BL31ID is one of the four 30 m-long straight sections, where a horizontal divergence of 
the electron beam is significantly reduced to 12 $\mu$rad, as shown in Fig.~\ref{fig:bl31}. Therefore, 
a laser-backscattering photon beam is still collimated even after a long propagation, and a large detector system 
can be constructed at an open space outside the storage ring building. In addition, it is expected that the better 
divergence improves a photon energy resolution at a tagger measurement. Compared with the LEPS experiment, the 
new beamline experiment is characterized by the following two features: (1) The photon beam intensity will be 
increased by about one order for both types of photon beams produced from UV and deep UV lasers. For this purpose, 
a large aperture beamline is being constructed to achieve a simultaneous injection of four lasers. An optical 
handling of a laser cross section to increase an energy density is also under considerations as discussed later. 
High statistics data of hadron photoproductions, including the reactions whose production rates are relatively low,
will be available to investigate exotic hadron structures, hadron-hadron correlations, hadron mass generation,
and so on. (2) A large detector system, which covers $\sim$4$\pi$ solid angles, is being constructed based on 
a 1 Tesla solenoidal magnet with a bore diameter of 3 m and a length of 2.2 m. The magnet and a detector system 
have been transferred from the E787/E949 experiment, which studied rare kaon decays using stopped K$^+$s at BNL. 
The detector system will be modified to fit with a hadron experiment and to achieve a momentum resolution of 
$\sim$1\%. Systematic measurements of differential cross sections and polarization observables depending on a 
production angle will become possible to investigate hadron natures stepping into details of reaction mechanisms.

  At the new beamline, it is planned to inject laser lights from a side-wall of concrete shields and reflect them 
toward a collision point at a beamline chamber close to the straight section by setting an aluminum-coated silicon 
mirror. This mirror will be modified to have a thin slit to escape x-rays and a small hole to pass the photon beam. 
Adoption of the side-way injection is favorable to shorten the laser paths, whose cross sections are expanded to make 
a focus, and reduce the aperture of beamline chambers, where an ultrahigh vacuum is necessary. Moreover, radiation 
shields to surround the laser system can be minimized without a laser hutch. Inside a clean room beside the concrete 
wall, four mode-lock UV lasers (355 nm, 8 W) and four solid-state deep UV lasers (266 nm, 1 W) will be separately 
placed with beam expanders, reflection mirrors, and half-wave plates. In order to achieve a four laser injection, 
each two laser paths will be first merged at a prism, and the two merged paths will be simultaneously injected with 
a slight difference of two prism heights. Based on the successful results of two laser injection at the LEPS experiment, 
an increase of photon beam intensity relative to that of a single laser injection is expected to be at most a factor 
of 4, although there should be a reduction of the gain due to finite injection angles. In addition to the efforts of
multi-laser injection, progresses of laser technologies themselves are fast producing higher power CW (or quasi-CW) 
lasers. A UV power output of 16 W at $\lambda$=355 nm is already available in the same series of commercial products 
as the 8 W mode-lock laser. Although a diameter of the 16 W laser is increased to 1.4 mm, the beam expander for the 
8 W laser has been designed to be usable also for the new laser. The 16 W laser has been tested at the LEPS facility, 
resulting in twice of the photon beam intensity compared with that by the 8 W laser. At the tests, a few problems 
were found for a long-term use of the 16 W laser; Firstly, a pointing stability of the laser beam was not good to 
collide it with a narrow electron beam at the 37 m-downstream focus point. However, this problem has been fixed by 
introducing a higher cooling power system for the laser. Furthermore, serious damages of the aluminum-coated silicon 
mirrors and the entrance window were observed with a long-term exposure of the 16 W laser. This has been also solved 
at the tests by adding NEG pumps around these optical components and improving a vacuum level. In case of replacing 
all of the four 8 W lasers to the 16 W lasers, a total gain of the photon beam intensity will be surely increased
twice.

  As mentioned in the previous section, a cross section of the electron beam is horizontally wide because of
synchrotron radiation. In contrast, a laser beam possesses a round shape because its transverse emission mode 
usually corresponds to TEM$_{00}$. A spacial overlap of the two beams is not sufficient, and transforming either 
of beam shape to the other's is effective to increase their collision rate. It is possible to change the electron 
beam shape to a circle, but many magnets are necessary to be installed at the long straight section. This option 
is not realistic because its cost is too expensive and a collision point becomes farther in spite of the limitations 
from beamline chamber apertures. An optical shaping of the laser beam is rather easier to produce an elliptical 
cross section. Here a larger expansion of the laser beam makes a smaller beam waist size at the focus point, as 
recognized from Eqn.~\ref{eq:laserprop}. Therefore, a small beam expander, which expands a laser diameter twice 
only in one direction with cylindrical concave and convex lenses (cylindrical expander), has been produced so that 
it can be mounted just in front of a normal beam expander, which isotropically expands the input elliptical beam. 
An output lens diameter of this normal expander is enlarged to 80 mm, while the magnification factor is kept to 
be 28.6, which is the same as that of the beam expander described in the previous section. A UV laser beam shape 
has been examined with the cylindrical expander and the $\phi$80 mm normal expander by monitoring a focus cross 
section 10 m downstream using a laser beam profiler, which covers a 20 mm-square area by a CMOS pixel sensor. 
In comparison with the case using only the normal expander, it has been confirmed that only a vertical size is 
reduced about half, producing an elliptical beam as designed. Importantly, a peak energy density has increased 
nearly twice with the cylindrical expander. However, it is expected that a problem arises from the decrease of 
laser beam waist length (Rayleigh range), which typically defines an effective region of backward Compton 
scatterings. The technique of beam shape transformation will be useful when the e$\gamma$ collision range is 
effectively short, for example, with a finite injection angle. Additionally, a technique of a long range 
non-diffractive beam (LRNB) \cite{lrnb} is under considerations to extend the beam waist length with a modest 
reduction of the peak energy density. The aperture of the LEPS2 beamline has been designed by taking into account 
a possibility of the laser injection with an expanded elliptical cross section.

  The construction of the LEPS2 beamline including the laser injection system will continue until 2012. 
The BNL-E787/E949 detectors and the solenoidal magnet were transferred to a new experimental building at SPring-8 
in late 2011. A new detector system will be prepared in parallel with the beamline construction. A beam commissioning 
with a simple detector setup will be done in 2012, and an experiment with the full setup of laser injection system 
and detectors is planned to start in 2013. Physics results with higher precisions and more systematic coverages,
compared with the LEPS experiment, will be available to obtain new insights of hadron natures as discussed above.

  Further upgrades of photon beam energy and intensity will mainly depend on availabilities of a high energy and 
high current electron storage ring and progresses of laser technologies with a short wavelength and a high power. 
High mass particles like glueballs and hybrids may be analyzed with such an upgraded photon beam.
In lattice QCD, lowest mass candidates of scaler and tensor glueballs are predicted at $\sim$1.7 GeV/c$^2$ and 
$\sim$2.4 GeV/c$^2$, respectively, and they are produced if a photon energy exceeds 3.2 GeV and 5.5 GeV.
Near-threshold photoproduction of a vector charmonium, J/$\psi$, becomes also possible above a photon energy of
8.2 GeV. Currently, it is difficult to produce a CW or quasi-CW laser with a wavelength shorter than $\sim$250 
nm, while a deep UV output power at $\lambda$ $=$ 266 nm may be able to be raised up with a better SHG crystal.
Instead, it may be worth considering a re-injection of an x-ray from an insertion device into the same 
storage ring by reflecting it back with multi-layer mirrors \cite{x-inj}. For example, an x-ray with a 100 eV or 
more energy results in a maximum photon energy above 7.4 GeV by backward Compton scattering at SPring-8. Future 
developments to produce the GeV photon beam, mentioned above, is expected to expand a possible physics range.

\section*{Acknowledgements}
The author thanks the members of the LEPS Collaboration and the staff at SPring-8 for supporting
developments of the Laser Compton Scattering beam at BL33LEP and BL31LEP. The author gratitudes for
T. Aruga, who gave many useful comments about the laser beam propagation, and the technical staffs 
of Oxide Corp. and Coherent Inc., who were helpful for the laser operations at SPring-8. Research
programs of the LEPS experiment were supported in part by the Ministry of Education, Science, Sports 
and Culture of Japan, the National Science Council of Republic of China (Taiwan), and the KOSEF of 
Republic of Korea.

%

\end{document}